\documentclass[a4paper,11pt]{article}

\usepackage{jheppub} 
\usepackage{graphicx}
\usepackage{color}
\usepackage{hyperref}
\usepackage{dsfont}
\usepackage{here}
\usepackage[section]{placeins}
\usepackage{subcaption}
\usepackage{array}
\usepackage{multirow}
\usepackage{tikz,xcolor}
\definecolor{lime}{HTML}{A6CE39}
\DeclareRobustCommand{\orcidicon}{%
	\begin{tikzpicture}
	\draw[lime, fill=lime] (0,0) 
	circle [radius=0.16] 
	node[white] {{\fontfamily{qag}\selectfont \tiny ID}};	\draw[white, fill=white] (-0.0625,0.095) 
	circle [radius=0.007];	\end{tikzpicture}
	\hspace{-2mm}}
\foreach \x in {A,...,Z}{%
	\expandafter\xdef\csname orcid\x\endcsname{\noexpand\href{https://orcid.org/\csname orcidauthor\x\endcsname}{\noexpand\orcidicon}}
}

\begin{document}

\title{Grassmann tensor renormalization group approach to $(1+1)$-dimensional two-color lattice QCD at finite density}

\author[a]{Kwok Ho Pai\orcidA{}\,}
    \affiliation[a]{Department of Physics, The University of Tokyo, Bunkyo-ku, Tokyo 113-0033, Japan}
    \emailAdd{hopai.kwok@phys.s.u-tokyo.ac.jp}
    
\author[b,a]{Shinichiro Akiyama\orcidB{}\,}
    \affiliation[b]{Center for Computational Sciences, University of Tsukuba, Tsukuba, Ibaraki 305-8577, Japan}
    \emailAdd{akiyama@ccs.tsukuba.ac.jp}
    
\author[a,c,d]{Synge Todo\orcidC{}}
    \affiliation[c]{Institute for Physics of Intelligence, The University of Tokyo, Bunkyo-ku, Tokyo 113-0033, Japan}
    \affiliation[d]{Institute for Solid State Physics, The University of Tokyo, Kashiwa, Chiba 277-8581, Japan}
    \emailAdd{wistaria@phys.s.u-tokyo.ac.jp}

%\date{March 2024}
\preprint{UTCCS-P-158}

\abstract{
We construct a Grassmann tensor network representing the partition function of (1+1)-dimensional two-color QCD with staggered fermions. The Grassmann path integral is rewritten as the trace of a Grassmann tensor network by introducing two-component auxiliary Grassmann fields on every edge of the lattice.
We introduce an efficient initial tensor compression scheme to reduce the size of initial tensors. 
The Grassmann bond-weighted tensor renormalization group approach is adopted to evaluate the quark number density, fermion condensate, and diquark condensate at different gauge couplings as a function of the chemical potential. Different transition behavior is observed as the quark mass is varied. We discuss the efficiency of our initial tensor compression scheme and the future application toward the corresponding higher-dimensional models.
}

\maketitle

\section{Introduction}

Since numerical simulations of lattice field theories based on the Monte Carlo approach generally suffer from the sign problem, the tensor renormalization group (TRG)~\cite{Levin:2006jai}, which is a variant of real-space renormalization group on a tensor network and does not rely on any sampling methods, has become an attractive strategy for the Lagrangian formulation. 
Improvements in two dimensions~\cite{Xie:2009zzd,PhysRevLett.115.180405,PhysRevLett.118.110504,PhysRevE.97.033310,PhysRevB.105.L060402}, efficient higher dimensional algorithms~\cite{PhysRevB.86.045139,Adachi:2019paf,Kadoh:2019kqk,Yamashita:2021yxs,Nakayama:2023ytr}, and extensions to fermion systems~\cite{Gu:2010yh,Gu:2013gba,Shimizu:2014uva,Akiyama:2020sfo} have made the TRG approach more accessible to computational lattice field theories. 
During the last decade, TRG has been used to simulate various models not only in two dimensions but also in the higher dimensions such as the $\phi^4$ theory~\cite{Shimizu:2012zza,Shimizu:2012wfa,Kadoh:2018tis,Kadoh:2019ube,Delcamp:2020hzo,Akiyama:2020ntf,Akiyama:2021zhf}, the Schwinger model~\cite{Shimizu:2014uva,Shimizu:2014fsa,Shimizu:2017onf,Butt:2019uul,Yosprakob:2023tyr}, the Gross-Neveu model~\cite{Takeda:2014vwa,Akiyama:2023lvr,Akiyama:2020soe}, the gauge-Higgs model~\cite{Unmuth-Yockey:2018ugm,Bazavov:2019qih,Akiyama:2022eip,Akiyama:2023hvt}, and the pure Yang-Mills theory~\cite{Fukuma:2021cni,Hirasawa:2021qvh,Kuwahara:2022ubg,Yosprakob:2024sfd}.
Toward the TRG study of the quantum chromodynamics (QCD) at finite temperature and density, it is necessary to develop a methodology to deal with the non-Abelian gauge theory in the presence of dynamical fermions.
Recently, several attempts have been made for the two-dimensional QCD in the infinite-coupling limit~\cite{Bloch:2022vqz} and two-color QCD with reduced staggered fermions~\cite{Asaduzzaman:2023pyz}.
There are also various previous studies of the (1+1)-dimensional QCD based on the Hamiltonian formalism employing the variational tensor network method such as matrix product state (MPS)~\cite{Kuhn:2015zqa,Silvi:2016cas,Banuls:2017ena,Sala:2018dui,Silvi:2019wnf,Rigobello:2023ype,Liu:2023lsr,Hayata:2023pkw}.
Although the TRG and MPS approaches are different tensor network methods, they share several aspects in their numerical calculations. 
In particular, both approaches require the discretization for the non-Abelian gauge fields.
Therefore, findings from Lagrangian-based calculations may also yield useful information for the Hamiltonian-based approach, and vice versa.

Two-color QCD, the $SU(2)$ Yang-Mills theory coupled with fermions in the fundamental representation of the gauge group, serves as a nice primary target for the tensor network approach. 
It has fewer degrees of freedom than the three-color QCD and the validity of the numerical results can be verified through comparison with the Monte Carlo method, particularly in four dimensions; two-color QCD has been investigated as an alternative to the three-color QCD because its fermion determinant is positive when the number of quark flavors is even. 
This enables Monte Carlo simulations at finite density in contrast to the three-color case~\cite{Kogut:1983ia,NAKAMURA1984391}. 
Two-color QCD shares several properties with three-color QCD such as the spontaneous breaking of chiral symmetry at low density with a restoration of the symmetry at larger density. 
In addition, there is an instability of the Fermi sphere against the formation of diquark condensate at sufficiently large density in both theories. 
Although diquarks in two-color QCD are color singlets, which differs from the case in three-color QCD, lattice simulations of two-color QCD are expected to provide hints about the nature of deconfined phase and the mechanism of condensate formation in dense three-color QCD.

In this work, we construct a Grassmann tensor network representation for the partition function of (1+1)-dimensional two-color QCD with staggered fermions and apply the Grassmann bond-weighted TRG (BTRG) algorithm~\cite{PhysRevB.105.L060402,Akiyama:2022pse} to evaluate the expectation values of several physical observables.
As an advantage of the Grassmann tensor network formulation, we can directly deal with the fermionic degrees of freedom without introducing pseudo-fermion.
We focus on the finite density regime in the thermodynamic and vanishing temperature limits and investigate whether the TRG computation can capture the effect of finite gauge coupling, quark mass, and diquark source term.
For the four-dimensional two-color QCD, the phase diagram in the $T$-$\mu$-$m$ space, with the temperature $T$, chemical potential $\mu$, and quark mass $m$, has been intensively studied~\cite{Hands:1999md,Kogut:1999iv,Kogut:2002cm,Nishida:2003uj,Hands:2007uc,Strodthoff:2011tz,Cotter:2012mb,Braguta:2016cpw,Iida:2019rah,Iida:2024irv}.
With a view to future applications of the TRG approach to the four-dimensional theory, we particularly compute the number density, chiral condensate, and diquark condensate, commonly used to investigate the phase structure.
In this study, the chiral and diquark condensates are evaluated by explicitly breaking the $U(1)_{A}$ and $U(1)_{V}$ symmetries because continuous symmetry is not spontaneously broken in two dimensions~\cite{Mermin:1966fe,Coleman:1973ci}.
We confirm that their behavior is consistent with each other.
We also see that the number density does not saturate in regions of larger chemical potential as the gauge interaction is weakened, approaching the continuum limit, in other words.
Our work is the first TRG study of the non-Abelian gauge theory with finite gauge coupling in the presence of the fermions at finite density.

This paper is organized as follows. 
In section 2, we introduce the tensor network representation of the lattice theory. 
In section 3, we briefly review BTRG and describe the initial tensor compression scheme facilitating practical calculations. 
In section 4, we present numerical results in the infinite coupling limit and the finite coupling regime. 
Section 5 is devoted to the conclusion and outlook.

\section{Tensor network representation of $SU(N)$ Yang-Mills theory with staggered fermion}

\subsection{Lattice theory}
We consider a $(1+1)$-dimensional $SU(N)$ Yang-Mills theory coupled with staggered fermion defined on a square lattice $\Lambda$ with volume $V=L^2$. The action is given by
\begin{align} 
\label{eq:S}
    S = S_f + S_g
\end{align}
with
\begin{align} 
\label{eq:S_f}
    S_f 
    &= \sum_{n\in\Lambda, \, \nu=1,2} \frac{p_\nu(n)}{2}\Big[ {\rm e}^{\mu \delta_{\nu, 2}} \bar{\chi}(n) U_\nu(n) \chi(n+\hat{\nu}) - {\rm e}^{-\mu \delta_{\nu, 2}} \bar{\chi}(n+\hat{\nu}) U^\dagger_\nu(n) \chi(n) \Big] \nonumber\\
    &+ m \sum_{n}\bar{\chi}(n) \chi(n) ,
\end{align}
\begin{align} 
\label{eq:S_g}
    S_g = -\frac{\beta}{N} \sum_{n} \text{Re} \, \text{Tr}\, U_1(n) U_2(n+\hat{1}) U_1^\dagger(n+\hat{2}) U_2^\dagger(n),
\end{align}
where each lattice site is labeled by $n=(n_1, n_2)$ with $n_{\nu}=1,\cdots,L$. 
The lattice spacing $a$ is always set to $a=1$.
The $N$-component staggered fermions are denoted by $\chi(n)$ and $\bar{\chi}(n)$ with the mass $m$ and chemical potential $\mu$.
We define the staggered phase function as $p_1(n)=1$ and $p_2(n)=(-1)^{n_1}$.
The link variables $U_{\nu}(n) \in SU(N)$ live on the link from the site $n$ to $n+\hat{\nu}$ and the inverse gauge coupling is represented by $\beta$.
The partition function is given by a Euclidean path integral
\begin{align} 
\label{eq:Z}
    Z = \int \mathcal{D}U \mathcal{D}\chi \mathcal{D}\bar{\chi}\,\text{e}^{-S},
\end{align}
where $\mathcal{D}U \equiv \prod_n {\rm d}U_1(n)\,{\rm d}U_2(n)$ and ${\rm d}U$ is the Haar measure of $SU(N)$. 
The Grassmann path integral measure is defined by $\mathcal{D}\chi \mathcal{D}\bar{\chi} \equiv \prod_n \prod_{c=1}^{N} \mathrm{d}\chi_c(n) \mathrm{d}\bar{\chi}_c(n)$.
We always assume the periodic boundary conditions for link variables and the (anti-)periodic boundary conditions for staggered fermions in the $\hat{1}$ ($\hat{2}$) direction.

\subsection{Tensor network representation}
\begin{figure}[htbp]
    \centering
    \includegraphics[scale=0.62]{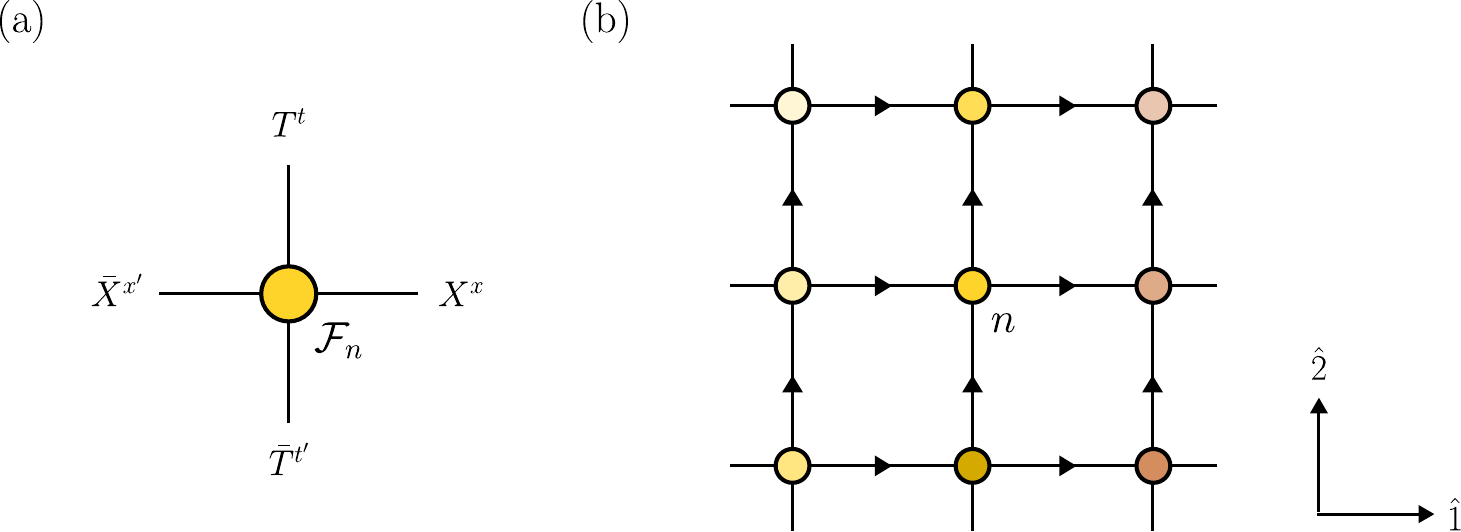}
    \caption{(a) Graphical representation of $\mathcal{F}_{n}$ and the auxiliary Grassmann fields that the local tensor carries. 
    (b) Tensor network representation of $Z_f[U]$ in Eq.~\eqref{eq:zf}.
    Notice that every local tensor is distinct from each other because $\mathcal{F}_{n}$ depends on the $U_\nu$ residing on the links connected to it.}
    \label{fig:TF}
\end{figure}

We begin with considering the fermionic sector.
Expressing Eq.~\eqref{eq:Z} as
\begin{align}
\label{eq:z_full}
    Z=\int \mathcal{D}U Z_{f}[U]~\text{e}^{-S_{g}},
\end{align}
where
\begin{align}
\label{eq:zf}
    Z_{f}[U]
    =
    \int \mathcal{D}\chi \mathcal{D}\bar{\chi}~\text{e}^{-S_{f}},
\end{align}
we firstly derive a tensor network representation of $Z_{f}[U]$.
Following the formalism in Ref.~\cite{Akiyama:2020sfo}, we introduce two $N$-component auxiliary Grassmann fields $\eta_{\nu}(n)$ and $\zeta_{\nu}(n)$ on every link of the lattice. 
We can decompose the fermion hopping terms in Eq.~\eqref{eq:S_f} with these auxiliary Grassmann fields via
\begin{align}
\label{eq:hop_forward}
    &\exp\left[ -\frac{p_\nu(n)}{2} \text{e}^{\mu \delta_{\nu, 2}}  \bar{\chi} (n) U_\nu(n) \chi(n+\hat{\nu}) \right] \nonumber\\
    &= 
    \int_{\bar{\eta}_{\nu}(n),\eta_{\nu}(n)}
    \exp\left[ 
        -\bar{\chi}(n) \eta_{\nu}(n) 
        + \frac{p_\nu(n)}{2} \text{e}^{\mu \delta_{\nu, 2}} \bar{\eta}_{\nu}(n)  
        U_\nu(n)\chi(n+\hat{\nu})
    \right],
\end{align}
\begin{align}
\label{eq:hop_backward}
    &\exp\left[ \frac{p_\nu(n)}{2} \text{e}^{-\mu \delta_{\nu, 2}}  \bar{\chi} (n+\hat{\nu}) U^\dagger_\nu(n) \chi(n) \right] \nonumber\\
    &= \int_{\bar{\zeta}_{\nu}(n),\zeta_{\nu}(n)} 
    \exp\left[ 
        \bar{\chi}(n+\hat{\nu})U^\dagger_\nu(n)\bar{\zeta}_{\nu}(n) 
        - \frac{p_\nu(n)}{2} \text{e}^{-\mu \delta_{\nu, 2}} \zeta_{\nu}(n) \chi(n) 
    \right].
\end{align}
In Eqs.~\eqref{eq:hop_forward} and \eqref{eq:hop_backward}, we introduced the following short-hand notation,
\begin{align}
    \int_{\bar{\theta},\theta}
    =
    \prod_{c=1}^{N}\int{\rm d}\bar{\theta}_{c}{\rm d}\theta_{c}
    {\rm e}^{-\bar{\theta}_{c}\theta_{c}},
\end{align}
with the $N$-component Grassmann variables $\theta$ and $\bar{\theta}$.
By these decompositions, it is easy to integrate out the original Grassmann fields $\chi(n)$ and $\bar{\chi}(n)$ at each lattice site independently.
The integration at the site $n$ results in a Grassmann tensor $\mathcal{F}_{n}$ such as
\begin{align} 
\label{eq:Tf2}
    \mathcal{F}_{n}[U] 
    = 
    \sum_{x,t,x',t'}
    (F_{n})_{xtx't'}[U_{1}(n-\hat{1}),U_{2}(n-\hat{2})]
    X^{x}T^{t}\bar{X}^{x'}\bar{T}^{t'}.
\end{align}
Due to Eqs.~\eqref{eq:hop_forward} and \eqref{eq:hop_backward}, the coefficient tensor $(F_{n})_{xtx't'}$ depends on the link variables $U_{1}(n-\hat{1})$ and $U_{2}(n-\hat{2})$ and so does $\mathcal{F}_{n}$.
Since we are dealing with the $N$-component fermion theory consisting of two types of hopping terms, the subscripts $x$, $t$, $x'$, $t'$ take their values on $\{0,1\}^{2N}$.
Therefore, the coefficient tensor $(F_{n})_{xtx't'}$ is a rank-$4$ complex-valued tensor whose bond dimension is $2^{2N}$. 
In the right-hand side of Eq.~\eqref{eq:Tf2}, $X^{x}$ abbreviates $\eta_{x,1}^{x_{1}}\cdots\eta_{x,N}^{x_{N}}\zeta_{x,1}^{x_{N+1}}\cdots\zeta_{x,N}^{x_{2N}}$ with $x=(x_{1},\cdots,x_{2N})$, as well as $T^{t}$.
Note that $X^{x}$ ($T^{t}$) resides on the positive $\hat{1}$ ($\hat{2}$) link connected to site $n$ as shown in Figure~\ref{fig:TF}~(a).
Similarly, $\bar{X}^{x'}=\bar{\zeta}_{x,N}^{x'_{2N}}\cdots\bar{\zeta}_{x,1}^{x'_{N+1}}\bar{\eta}_{x,N}^{x'_{N}}\cdots\bar{\eta}_{x,1}^{x'_{1}}$ with $x'=(x'_{1},\cdots,x'_{2N})$, as well as $\bar{T}^{t'}$, where $\bar{X}^{x'}$ ($\bar{T}^{t'}$) resides on the negative $\hat{1}$ ($\hat{2}$) link connected to site $n$.
See Appendix~\ref{app:elts_F} for the derivation of $F$.
We can restore $Z_{f}[U]$ in Eq.~\eqref{eq:zf} by contracting a Grassmann tensor network generated by $\mathcal{F}_{n}$:
\begin{align}
\label{eq:tn_z_f}
    Z_{f}[U]
    =
    {\rm gTr}
    %\prod_{n,\nu}
    %\int_{\bar{\eta}_{\nu}(n),\eta_{\nu}(n)}
    %\int_{\bar{\zeta}_{\nu}(n),\zeta_{\nu}(n)} 
    \left[
        \prod_{n}\mathcal{F}_{n}[U]
    \right]
    .
\end{align}
Here, ``gTr" denotes the integration over all auxiliary Grassmann fields.
The graphical representation of Eq.~\eqref{eq:tn_z_f} is illustrated in Figure~\ref{fig:TF}~(b).

\begin{figure}[htbp]
    \centering
    \includegraphics[scale=0.52]{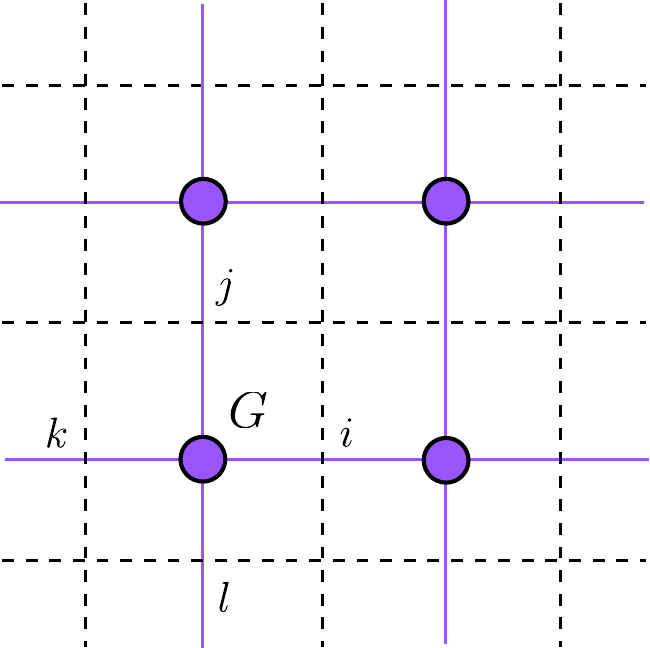}
    \caption{Tensor network representation for the partition function of pure $SU(N)$ Yang-Mills theory defined on a square lattice, which is indicated by the dashed lines.}
    \label{fig:TG}
\end{figure}

We now move on to the gauge sector with $\beta \neq 0$. 
We need to discretize the gauge group integration to derive the tensor network representation of Eq.~\eqref{eq:z_full}.
In this study, we use the method in Ref.~\cite{Fukuma:2021cni}, which approximates a group integration by an average of integrand evaluated using $K$ random $SU(N)$ matrices picked uniformly from the group manifold:
\begin{align} 
\label{eq:Umeasure}
    \int {\rm d}U\, f(U) 
    \simeq
    \frac{1}{K} \sum_{i=1}^{K} f(U_i).
\end{align}
We assign a tensor to each plaquette of the square lattice
\begin{align} 
\label{eq:tg}
    G_{i j k l} \equiv \frac{1}{K^2} {\rm e}^{(\beta / N) \operatorname{Re} \operatorname{Tr}\left(U_i U_j^{\dagger} U_k^{\dagger} U_l \right)},
\end{align}
where the indices $i,j,k,l$ range from $1$ to $K$. 
Each of them corresponds to one of the four links of the plaquette and its value specifies which random matrix in the set $\mathring{U}=\{U_1, U_2, \ldots, U_K\}$\footnote{In this study, the group integration of every link variable is approximated using the same set of random matrices $\mathring{U}$. One can use more than one set of matrices for different links of the lattice at the cost of increasing the number of distinct initial tensors.} 
is being substituted into the right-hand side of Eq.~\eqref{eq:tg}. 
The partition function of pure $SU(N)$ Yang-Mills theory defined on a square lattice is given by the trace of a tensor network composed of $G$, as shown in Figure~\ref{fig:TG}.

\begin{figure}[htbp]
    \centering
    \includegraphics[scale=0.55]{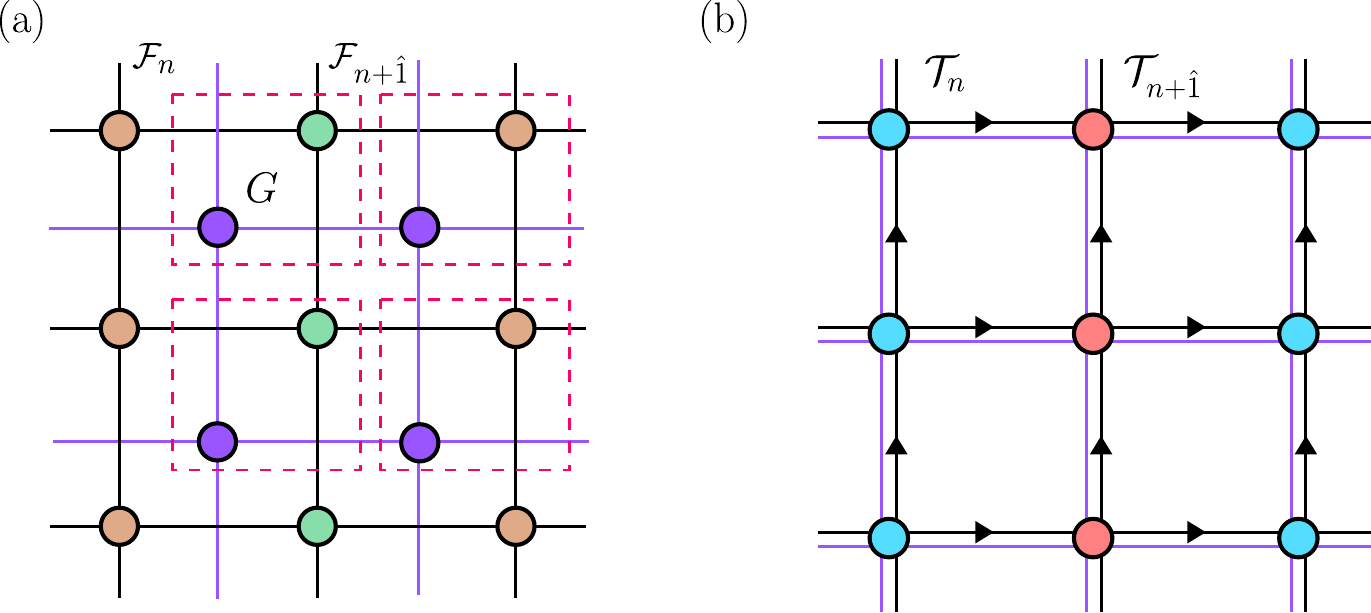}
    \caption{(a) $\mathcal{F}_{n}$ and $G$ are combined to form a new Grassmann tensor. 
    (b) Tensor network representation for the partition function of the full theory, which is composed of two initial tensors with bond dimension $2^{2N}K$.}
    \label{fig:T_merge}
\end{figure}

Now, we combine $\mathcal{F}_{n}$ and $G$ on the plaquette with $n$ being its top right corner to form the initial tensor (see Figure~\ref{fig:T_merge}~(a)):
\begin{align} 
\label{eq:totalT}
    \mathcal{T}_{n}
    = 
    \mathcal{F}_{n}\cdot G.
\end{align}

We then regard $\mathcal{T}_{n}$ as a new Grassmann tensor whose coefficient tensor is given by
\begin{align}
\label{eq:t_full}
    (T_{n})_{xtx't'}
    =
    (F_{n})_{x_{f}t_{f}x'_{f}t'_{f}}[t_{g},x_{g}]
    \cdot 
    G_{x_{g}t_{g}x'_{g}t'_{g}}
    ,
\end{align}
where the subscripts in $F_{n}$ are marked with $f$ and those in $G$ are marked with $g$.
In the left-hand side of Eq.~\eqref{eq:t_full}, we defined a super index $q$ by $q=(q_{f},q_{g})$ with $q=x,t,x',t'$.
Therefore, the bond dimension of $T_{n}$ is $2^{2N}K$.\footnote{
We comment on a Grassmann tensor network formulation of the $(1+1)$-dimensional $SU(2)$ Yang-Mills theory with the standard staggered fermion recently discussed in Ref.~\cite{Asaduzzaman:2023pyz}. 
Although their formulation results in the tensor network whose bond dimension is $2^{8}K$, our construction results in $2^{4}K$.
This is because our derivation introduces the $N$-component auxiliary fermions for forward and backward hopping terms as in Eqs.~\eqref{eq:hop_forward} and \eqref{eq:hop_backward}.
On the other hand, Ref.~\cite{Asaduzzaman:2023pyz} introduces the $N^{2}$-component auxiliary fermions for each hopping term.
}
The partition function in Eq.~\eqref{eq:z_full} is approximately expressed by using the fundamental tensor $\mathcal{T}_{n}$ as
\begin{align}
\label{eq:gtn_full}
    Z
    \simeq
    Z(K)
    =
    {\rm gTr}\left[
        \prod_{n}\mathcal{T}_{n}
    \right]
    %\sum\int\prod_{n}\mathcal{T}_{n}
    ,
\end{align}
as illustrated in Figure~\ref{fig:T_merge}~(b).
Note that ``gTr" in Eq.~\eqref{eq:gtn_full} means the summation over all subscripts marked by $g$ and the integration over all auxiliary Grassmann fields.

\begin{figure}[htbp]
    \centering
    \includegraphics[scale=0.52]{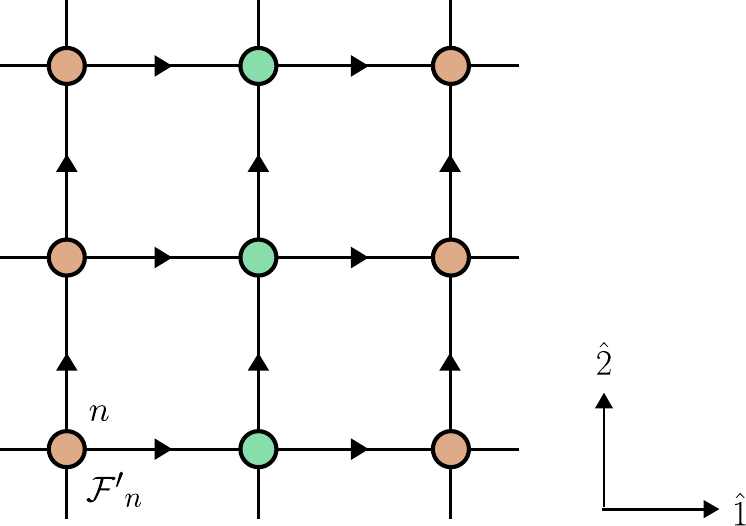}
    \caption{Tensor network representation of $Z_{\beta \to 0}$ in Eq.~\eqref{eq:Z_beta0}.}
    \label{fig:TF_inf}
\end{figure}

We finally remark that the infinite coupling limit $\beta \to 0$ can be easily taken within the current Grassmann tensor network formulation.
In this limit, we can perform the $SU(N)$ group integration exactly for all link variables on the lattice because the dependence of any $U_\nu(n)$ now appears in only one local tensor in our tensor network representation:
\begin{align}
\label{eq:Z_beta0}
    Z_{\beta\to0}
    =
    \int \mathcal{D}U Z_{f}[U]
    =
    \rm{gTr}
    %\prod_{n,\nu}
    %\int_{\bar{\eta}_{\nu}(n),\eta_{\nu}(n)}
    %\int_{\bar{\zeta}_{\nu}(n),\zeta_{\nu}(n)} 
    \left[
        \prod_{n}\mathcal{F}'_{n}
    \right]
    ,
\end{align}
where
\begin{align}
    \mathcal{F}'_{n}
    =
    \prod_{\nu}
    \int {\rm d}U_{\nu}
    \mathcal{F}_{n}[U]
    = 
    \sum_{x,t,x',t'}
    \left(
        \int {\rm d}U_{1}{\rm d}U_{2}
        (F_{n})_{xtx't'}[U_{1},U_{2}]
    \right)
    X^{x}T^{t}\bar{X}^{x'}\bar{T}^{t'}.
\end{align}
The bond dimension of $\mathcal{F}'_{n}$ is $2^{2N}$. 
The graphical representation of Eq.~\eqref{eq:Z_beta0} is shown in Figure~\ref{fig:TF_inf}.

\section{Algorithm}

\subsection{A quick review of BTRG}
The coarse-graining transformation of a tensor network can be facilitated using tensor renormalization group (TRG) algorithms. 
In TRG, the low-rank approximation of tensors, which is based on the singular value decomposition (SVD), is used to perform tensor contractions. 
In this study, the bond-weighted tensor renormalization group (BTRG) algorithm~\cite{PhysRevB.105.L060402} is employed. 
Bond weights, which are some power $k$ of the singular values from the SVD in the previous RG iteration, are introduced on the edges of the tensor network. 
It was suggested and confirmed in the case of the two-dimensional Ising model~\cite{PhysRevB.105.L060402} and massless free Wilson fermion~\cite{Akiyama:2022pse} that the optimal choice for the hyperparameter $k$ is $-1/2$ for square tensor networks. 
The hyperparameter $k$ is always set to be $-1/2$ in this study, and $D$ is the bond dimension cutoff of the BTRG algorithm, which usually depends on the initial bond dimension.

\subsection{Initial tensor compression}
As we mentioned, the bond dimension of the initial tensors in the Grassmann tensor network representing the partition function of the full theory is $2^{2N}K$. 
Although we only investigate the two-color ($N=2$) case numerically in this study, the initial bond dimension is $16K$.
This implies that a very large $D$ is inevitable for accurate TRG results.

\begin{figure}[htbp]
    \centering
    \includegraphics[scale=0.55]{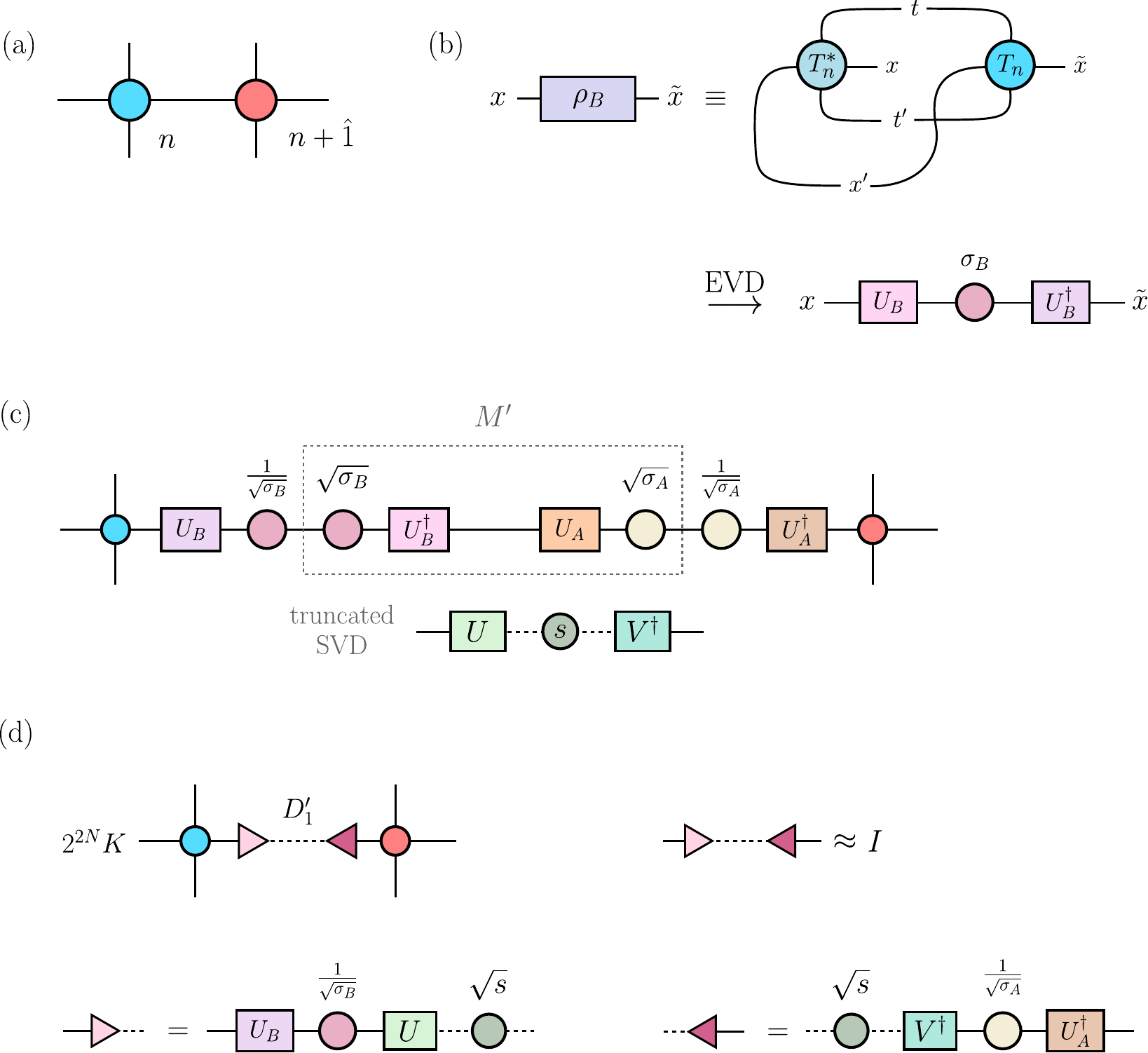}
    \caption{
    Procedure of constructing a pair of squeezers.
    (a) Two adjacent Grassmann tensors $\mathcal{T}_{n}$ and $\mathcal{T}_{n+\hat{1}}$.
    (b) From the Grassmann tensor $\mathcal{T}_{n}$, we define a Hermitian matrix $\rho_{B}$, whose EVD gives us a unitary matrix $U_{B}$ and the corresponding eigenvalue $\sigma_{B}$.
    We repeat the same procedure for $\mathcal{T}_{n+\hat{1}}$ and obtain a unitary matrix $U_{A}$ and the corresponding eigenvalue $\sigma_{A}$.
    (c) Inserting two pairs of invertible matrices, the truncated SVD is performed.
    (d) The truncated SVD in (c) defines the pair of squeezers.
    }
    \label{fig:squeezer}
\end{figure}

To tackle this problem, we propose an efficient tensor compression scheme that aims to find an accurate low-rank approximation for the initial tensors. 
The central idea is to insert a pair of squeezers, which approximates the identity, on every bond of the tensor network. 
By contracting an initial tensor with the squeezers connected to it, its bond dimension can be reduced. 
Now, we explain how to construct the two squeezers on a bond connecting $\mathcal{T}_{n+\hat{1}}$ and $\mathcal{T}_{n}$.
The procedure is graphically summarized in Figure~\ref{fig:squeezer}.

We first define a matrix notation for the coefficient tensors $T_{n}$ and $T_{n+\hat{1}}$ as
\begin{equation} 
\label{eq:mat_BA}
\begin{split}
&(M_A)_{x' (x t t')} = (T_{n+\hat{1}})_{x t x' t'} \, (-1)^{f_{x'}(f_x+f_t)} \, , \\
&(M_B)_{(t x' t') x} = (T_{n})_{x t x' t'} \, (-1)^{f_{x}(f_t+f_{x'}+f_{t'})} \, .
\end{split}
\end{equation}
In Eq.~(\ref{eq:mat_BA}), we introduced a Grassmann parity function $f_q$ for a super index $q=(q_f, q_g)$ as
\begin{equation} 
\label{eq:parity}
f_q = \sum_{i=1}^{2N} q_{f, i} \;\; \mathrm{mod} \; 2, 
\end{equation}
which diagnoses the Grassmann parity of $Q^{q_f}$ ($Q^{q_f} = X^{x_f}, T^{t_f}, \bar{X}^{x'_f}, \bar{T}^{t'_f}$). Therefore, Grassmann fields will not appear explicitly in the following discussion, and the corresponding Grassmann algebra has already been encoded by the sign factors in Eq.~(\ref{eq:mat_BA}). Under this notation, the question becomes a low-rank approximation of the matrix $M_B  M_A$. An obvious solution is a truncated SVD. In this study, a more computationally economical approach is considered.

We insert an identity $I=L_B^{-1} L_B R_A R_A^{-1}$ in the middle of $M_B M_A$ and perform an SVD on $M'\equiv L_{B} R_A = UsV^\dagger$. The resulting expression $M_B L_B^{-1} U s V^\dagger R_A^{-1} M_A$ is a compact SVD of $M_B M_A$ when
\begin{equation} 
\label{eq:LB}
E^\dagger E = I \;\;\; \textrm{with} \;\;\; E = M_B L_B^{-1}
\end{equation}
and
\begin{equation} 
\label{eq:RA}
W^\dagger W = I \;\;\; \textrm{with} \;\;\; W^\dagger = R_A^{-1}  M_A.
\end{equation}
To find an $L_B$ satisfying Eq.~(\ref{eq:LB}), we define a Hermitian matrix $\rho_B = M_B^\dagger M_B$ which corresponds to the coefficient tensor of $\mathcal{T}^\dagger_{n} \mathcal{T}_{n}$. The eigenvalue decomposition (EVD) on $\rho_B$ gives a unitary matrix $U_B$, which diagonalizes $\rho_B$ and the corresponding eigenvalues $\sigma_B$. Then, $L_B^{-1}=U_B \frac{1}{\sqrt{\sigma_B}}$ and $L_B=\sqrt{\sigma_B}U_B^\dagger$.
Similarly, another Hermitian matrix $\rho_A= M_A M_A^\dagger$ which corresponds to the coefficient tensor of $\mathcal{T}_{n+\hat{1}} \mathcal{T}^\dagger_{n+\hat{1}}$ is constructed and we perform an EVD on $\rho_A$. Then, $R_A^{-1}=\frac{1}{\sqrt{\sigma_A}} U_A^\dagger$ satisfies Eq.~(\ref{eq:RA}) with $R_A=U_A\sqrt{\sigma_A}$.
To obtain a low-rank approximation of $M_B M_A$, we only keep the largest $D'$ singular values and vectors in the decomposition $M'=UsV^\dagger$. $D' \leq 2^{2N}K$ is the smallest integer satisfying the following condition
\begin{equation} 
\label{eq:ratio}
\frac{\sum_{y=1}^{D'} s^2_y}{\sum_{y=1}^{2^{2N}K} s^2_y} \geq r,
\end{equation}
where $r\leq1$ is a parameter of this compression scheme. The number of singular values retained from the even sector of $M$ is denoted as $D'_{\textrm{even}} \leq 2^{2N-1}K$.
We then define the two squeezers as $P_{B}=U_B \frac{1}{\sqrt{\sigma_B}} U \sqrt{s}$ and $Q_{A} = \sqrt{s} V^\dagger \frac{1}{\sqrt{\sigma_A}} U_A^\dagger$. Without truncation, $P_{B} Q_{A} = I$ . The size of $P_{B}$ and $Q_{A}$ after truncation are $2^{2N}K \times D'$ and $D' \times 2^{2N}K$ respectively. The matrix representation of the compressed tensors at site $n$ and $n+\hat{1}$ are $M_B P_{B}$ and $Q_{A} M_A$ respectively. One can then read off the corresponding coefficient tensors $T'_{n}$ and $T'_{n+\hat{1}}$. 

\begin{figure}[htbp]
    \centering
    \includegraphics[scale=0.7]{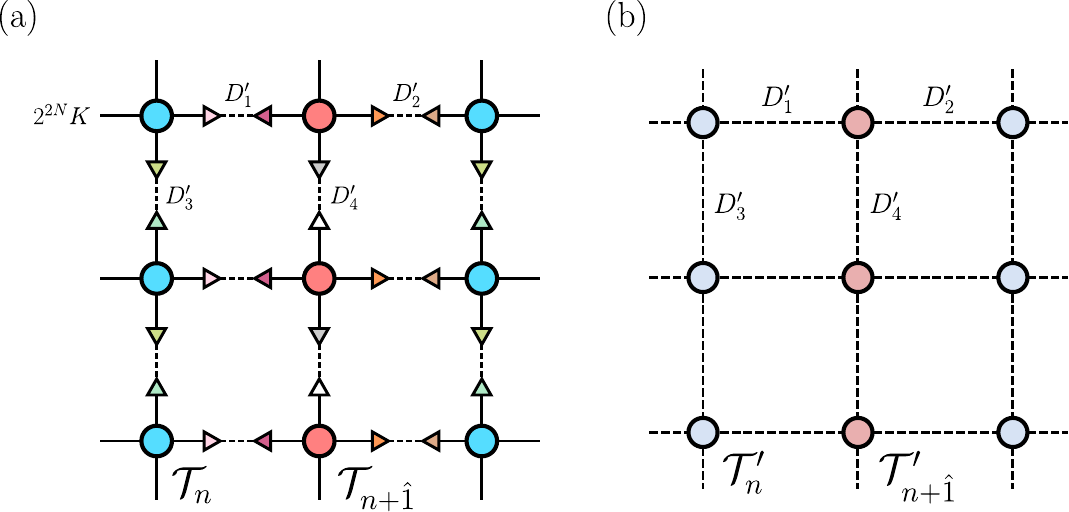}
    \caption{
    (a) Insertion of pairs of squeezers on every bond. 
    (b) Compressed Grassmann tensor network. 
    }
    \label{fig:T_squeezed}
\end{figure}

Through this procedure, the bond dimension being considered is reduced from $2^{2N}K$ to $D'$, which depends on the parameter $r$ according to Eq.~(\ref{eq:ratio}). As shown in Figure~\ref{fig:T_squeezed}, there are four different types of bonds in the Grassmann tensor network representing the partition function. One can repeat the above steps to construct a pair of squeezers for each of the remaining types of bonds and contract the initial tensor at each lattice with the four squeezers connected to it to obtain a compressed initial tensor.\footnote{The bond dimension of the four indices of the compressed initial tensor should be, in general, different from each other. It is because the bond dimension after compression $D'$ is determined by Eq.~(\ref{eq:ratio}), and the singular value spectrum $s$ is different for the four different types of bonds in the tensor network.}

\section{Numerical results}

By expressing the partition function (\ref{eq:Z}) in terms of the trace of a Grassmann tensor network, one can compute the free energy density $f=\textrm{ln}Z/V$ directly with TRG algorithms. Then, the expectation value of a physical observable $\langle O \rangle \equiv Z^{-1}\int \mathcal{D}U \mathcal{D}\chi \mathcal{D}\bar{\chi} \,O\, \textrm{e}^{-S}$ is given by the partial derivative of $f$.\footnote{
We note that the expectation value can also be expressed as the trace of a tensor network composed of an impurity tensor on some lattice sites and the earlier defined tensor $\mathcal{T}_{n}$ on the remaining sites~\cite{Yoshimura:2017jpk,MORITA201965}.
}
Two physical observables of interest in this study are the quark number density defined as
\begin{equation}
\label{eq: def_n}
\langle n \rangle = \frac{\partial f}{\partial \mu} \,\, ,
\end{equation}
and the fermion condensate, which is defined as
\begin{equation}
\label{eq: def_fc}
\langle \bar{\chi} \chi \rangle = \frac{\partial f}{\partial m} \,\, .
\end{equation}
In this study, the partial derivatives in Eqs.~(\ref{eq: def_n}) and (\ref{eq: def_fc}) are evaluated by the forward difference:
\begin{align}
\label{eq: finite_diff_number}
\langle n \rangle \simeq \frac{f(\mu+\Delta\mu) - f(\mu)}{\Delta\mu} \,\, , 
\end{align}
\begin{align}
\label{eq: finite_diff_fermion}
\langle \bar{\chi} \chi \rangle \simeq \frac{f(m+\Delta m) - f(m)}{\Delta m}.
\end{align}

In addition to the number density and fermion condensate, we also investigate the formation of diquark condensate. When we compute the diquark condensate, we add a diquark source term to the action (\ref{eq:S}) as
\begin{equation}
\label{eq: S_diquark}
S' = S + \frac{\lambda}{2} \sum_{n} \left[\chi^{T}(n) \sigma_2 \chi(n) + \bar{\chi}(n) \sigma_2 \bar{\chi}^T(n) \right] \,\, , 
\end{equation}
where $\lambda$ is a real parameter controlling the magnitude of the diquark source term, and the superscript $T$ means a transpose. The partition function and free energy density evaluated with $S'$ now depend on $\lambda$. It is straightforward to include this diquark source term in our Grassmann tensor network representation because it only contains single-site terms. See Appendix \ref{app:elts_F} for the derivation of the initial tensor elements with the new action $S'$. The expectation value of the diquark condensate defined as $\chi \chi \equiv  \sum_{n} \left(\chi^{T} \sigma_2 \chi + \bar{\chi} \sigma_2 \bar{\chi}^T\right)/2V$ can be computed by
\begin{equation}
\label{eq: diquark}
\langle \chi \chi \rangle \equiv \frac{1}{2V} \int \mathcal{D}U \mathcal{D}\chi \mathcal{D}\bar{\chi} \, \sum_{n} \left(\chi^{T} \sigma_2 \chi + \bar{\chi} \sigma_2 \bar{\chi}^T\right) \, \textrm{e}^{-S'} = \frac{\partial f}{\partial \lambda}.
\end{equation}

The two-color lattice QCD theory under consideration has the symmetry $U(1)_V \times U(1)_A$ at a finite chemical potential $\mu$, in the vanishing $\lambda$ limit and chiral limit $m=0$. 
The $U(1)_V$ and $U(1)_A$ symmetry correspond to the baryon number and axial charge conservation, respectively. 
A finite quark mass $m\neq0$ breaks the $U(1)_A$ symmetry: $\chi \to \mathrm{e}^{{\rm i}\alpha\epsilon(n)}\chi,\,\bar{\chi} \to \bar{\chi} \mathrm{e}^{{\rm i}\alpha\epsilon(n)}$~($\alpha\in\mathds{R}$) explicitly, and a finite $\lambda$ breaks the $U(1)_V$ symmetry: $\chi \to \mathrm{e}^{{\rm i}\alpha'}\chi,\,\bar{\chi} \to \bar{\chi} \mathrm{e}^{-{\rm i}\alpha'}$~($\alpha'\in\mathds{R}$), explicitly. 
Note that $\epsilon(n)$ is defined as $\epsilon(n)=(-1)^{n_{1}+n_{2}}$, which plays the similar role of $\gamma_{5}$ in the staggered fermion theory.
Therefore, these observables are intensively studied in the four-dimensional theory~\cite{Hands:1999md,Kogut:1999iv,Kogut:2002cm,Nishida:2003uj,Hands:2007uc,Strodthoff:2011tz,Cotter:2012mb,Braguta:2016cpw,Iida:2019rah,Iida:2024irv}.

However, since no spontaneous breaking of continuous symmetry happens in two dimensions~\cite{Mermin:1966fe,Coleman:1973ci}, we expect $\lim_{m\to0}\lim_{V\to\infty}\langle \bar{\chi} \chi \rangle = 0$, and $\lim_{\lambda\to0}\lim_{V\to\infty}\langle \chi \chi \rangle = 0$. 
To illustrate that the behavior of the aforementioned physical observables which reveals the phase structure of the theory in higher dimensions can be properly reproduced by the TRG approach, we always set $m>0$ and/or $\lambda>0$ in this study,\footnote{The quark number density $\langle n \rangle$ and the fermion condensate $\langle \bar{\chi} \chi \rangle$ are calculated at $\lambda=0$ in this study.} 
which explicitly breaks the symmetry of this model and allows a finite value of $\langle \bar{\chi} \chi \rangle$ and $\langle \chi \chi \rangle$. In particular, the diquark condensate is evaluated by
\begin{equation}
\label{eq: finite_diquark}
\langle \chi \chi \rangle \simeq\frac{f(\lambda+\Delta\lambda) - f(\lambda)}{\Delta\lambda}\,\, . 
\end{equation}

\subsection{Infinite coupling limit}

As mentioned in Section 2.2, the gauge group integration can be integrated exactly in the infinite coupling limit, and the bond dimension of the initial tensors is $2^{4}=16$. 
Therefore, the initial tensor compression scheme introduced in Section 3.2 is not employed at $\beta=0$.
In the following, we set a bond dimension cutoff $D=84$, which suffices to suppress the finite-$D$ effect.

\begin{figure}[htbp]
    \centering
    \includegraphics[scale=0.8]{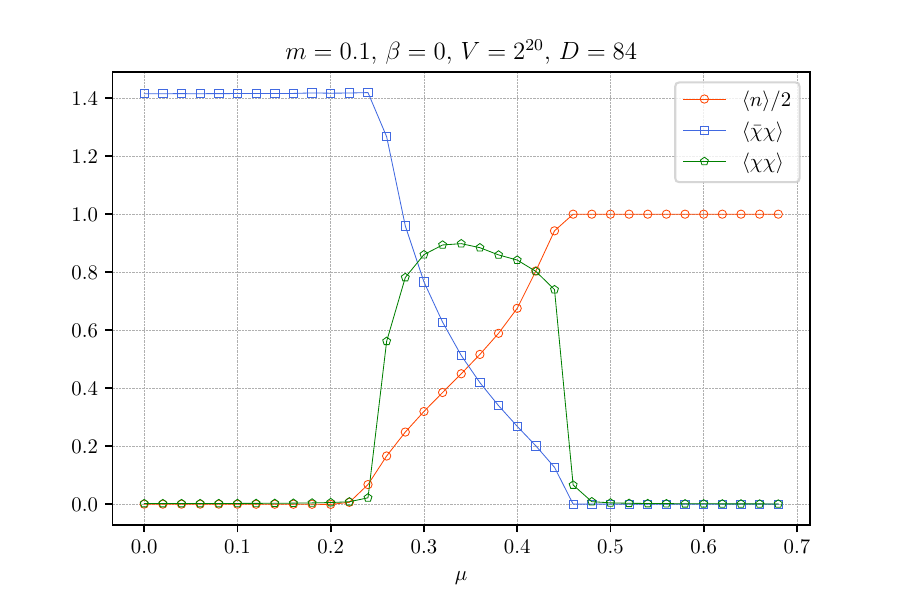}
    \caption{Quark number density $\langle n \rangle$, fermion condensate $\langle \bar{\chi} \chi \rangle$ and diquark condensate $\langle \chi \chi \rangle$ as a function of chemical potential $\mu$ at $m=0.1$, $\beta=0$, in the thermodynamic limit. The bond dimension in the calculations is $D=84$.
    To evaluate the numerical differences in Eqs.~\eqref{eq: finite_diff_number}, \eqref{eq: finite_diff_fermion}, and \eqref{eq: finite_diquark}, we set $\Delta \mu=0.04$, $\Delta m = 10^{-4}$, and $\lambda = \Delta\lambda = 10^{-4}$. 
    }
    \label{fig:m0.1_beta0}
\end{figure}

Figure \ref{fig:m0.1_beta0} shows the quark number density, fermion condensate, and diquark condensate with $m=0.1$ and on a lattice volume $V=2^{20}$. 
All these quantities do not depend on the chemical potential up to some point, $\mu_{c1}$.
These are the characteristic features of the so-called Silver-Blaze phenomena~\cite{Cohen:2003kd}.
Since the Silver-Blaze phenomena take place only in the thermodynamic and zero-temperature limits, the lattice volume $V=2^{20}$ is sufficiently large to obtain these limits.
We observe the Silver-Blaze region where $\langle n \rangle = 0$ from $\mu=0$ to $\mu_{c1} \approx 0.22$. 
As $\mu$ further increases, we see an intermediate phase, which extends over a finite region of chemical potential $\mu_{c1} < \mu < \mu_{c2}$, with $\mu_{c2} \approx 0.46$, characterized by $0<\langle n \rangle <2$ and non-zero $\langle \chi \chi \rangle$. 
As $\mu > \mu_{c2}$, $\langle n \rangle/2$ saturates to one, the maximum according to the Pauli exclusion principle on the lattice, and $\langle \chi \chi \rangle$ is suppressed to some values very close to zero.
On the other hand, the fermion condensate $\langle \bar{\chi} \chi \rangle$ takes a constant finite value in the Silver-Blaze region and decreases in the intermediate phase. 
When $\mu > \mu_{c2}$, $\langle \bar{\chi} \chi \rangle$ is reduced to some very small values as $\langle \chi \chi \rangle$. 
The qualitative behavior of the observables computed by BTRG at finite $m$ and/or $\lambda$ is similar to that reported in the previous mean-field theory~\cite{Nishida:2003uj}.

\begin{figure}[htbp]
    \centering
    \includegraphics[scale=0.8]{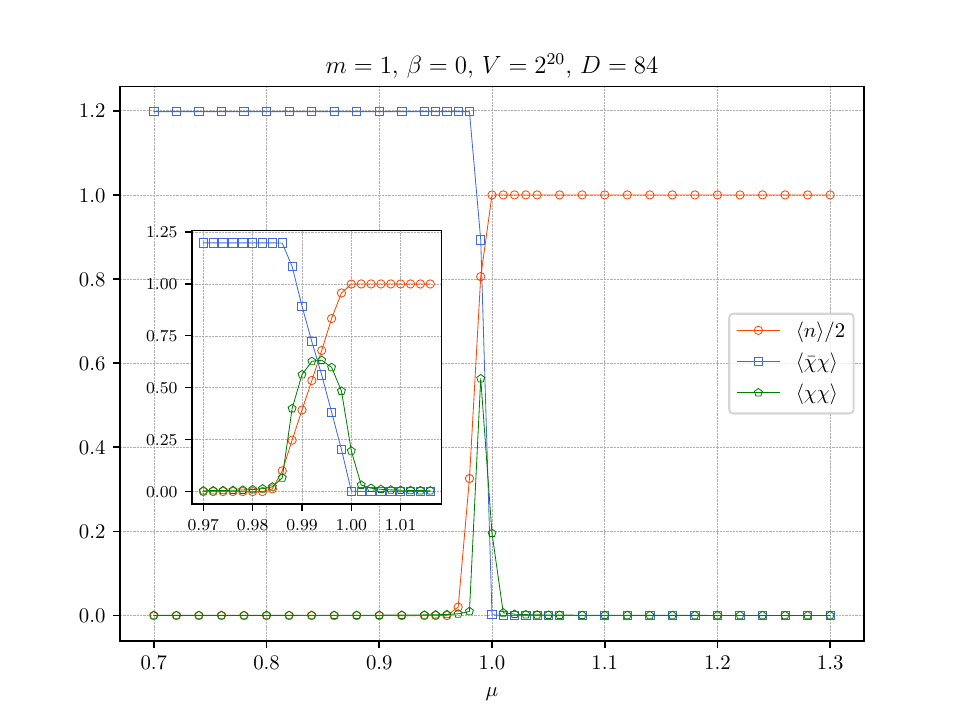}
    \caption{Quark number density $\langle n \rangle$, fermion condensate $\langle \bar{\chi} \chi \rangle$ and diquark condensate $\langle \chi \chi \rangle$ as a function of chemical potential $\mu$ at $m=1$, $\beta=0$, in the thermodynamic limit. The bond dimension in the calculations is $D=84$.
    To evaluate the numerical differences in Eqs.~\eqref{eq: finite_diff_number}, \eqref{eq: finite_diff_fermion}, and \eqref{eq: finite_diquark}, we set $\Delta \mu=0.02$, $\Delta m = 10^{-4}$, and $\lambda = \Delta\lambda = 10^{-4}$. The inset shows the three quantities in the intermediate phase, where $\langle n \rangle$ is evaluated using a finer $\Delta \mu = 0.004$.
    }
    \label{fig:m1_beta0}
\end{figure}

For a larger quark mass $m=1$ and the same lattice volume $V=2^{20}$, a sharp transition happens, and the intermediate phase becomes a very narrow region in $\mu$ as shown in Figure \ref{fig:m1_beta0}. 
We see $\mu_{c1} \approx 0.982$ and $\mu_{c2} \approx 1$. Apart from this difference, the qualitative behavior of $\langle n \rangle$, $\langle \bar{\chi} \chi \rangle$ and $\langle \chi \chi \rangle$ at $m=1$, $\beta=0$ is similar to that at $m=0.1$. 
Particularly, as shown in the inset of Figure \ref{fig:m1_beta0}, there is still no discontinuity of physical quantities observed for $m=1$ in the thermodynamic limit. It is fair to identify the transition in the intermediate phase as a crossover, instead of a first-order phase transition.

\begin{figure}[ht]
    \captionsetup[subfigure]{justification=centering}
        \subfloat[$\langle n \rangle$ as a function of $\mu$ at $m=0.1$.]{%
            \includegraphics[width=.5\linewidth]{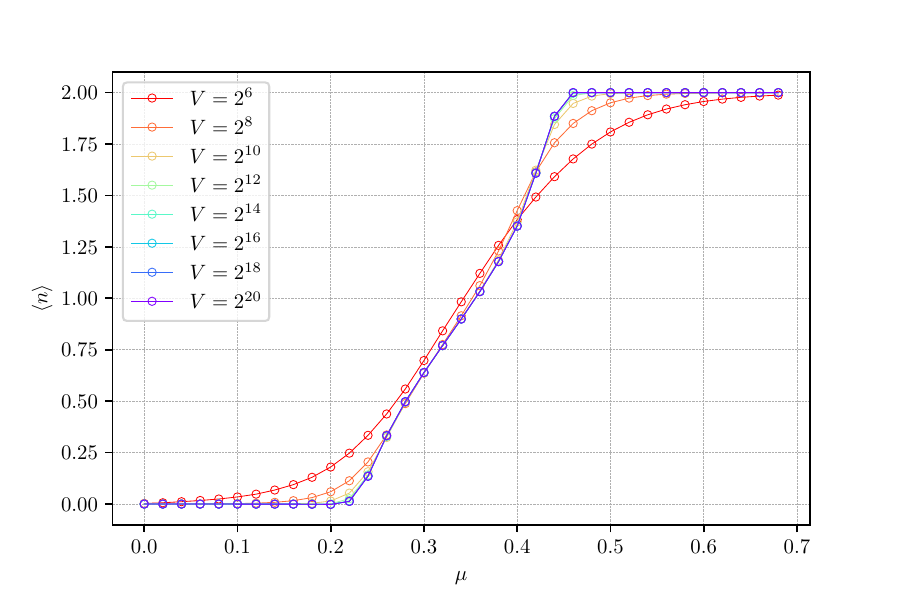}%
            \label{fig:number_m0.1_beta0_Vdependence}%
        }\hfill
        \subfloat[$\langle \chi \chi \rangle$ as a function of $\mu$ at $m=0.1$.]{%
            \includegraphics[width=.5\linewidth]{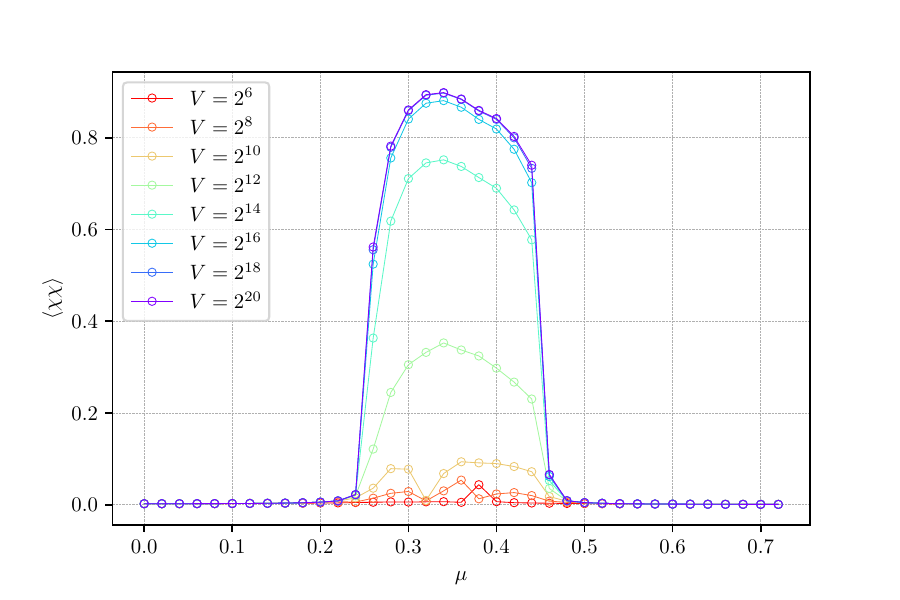}%
            \label{fig:diq_m0.1_beta0_Vdependence}%
        }\\
        \subfloat[$\langle n \rangle$ as a function of $\mu$ at $m=1$.]{%
            \includegraphics[width=.5\linewidth]{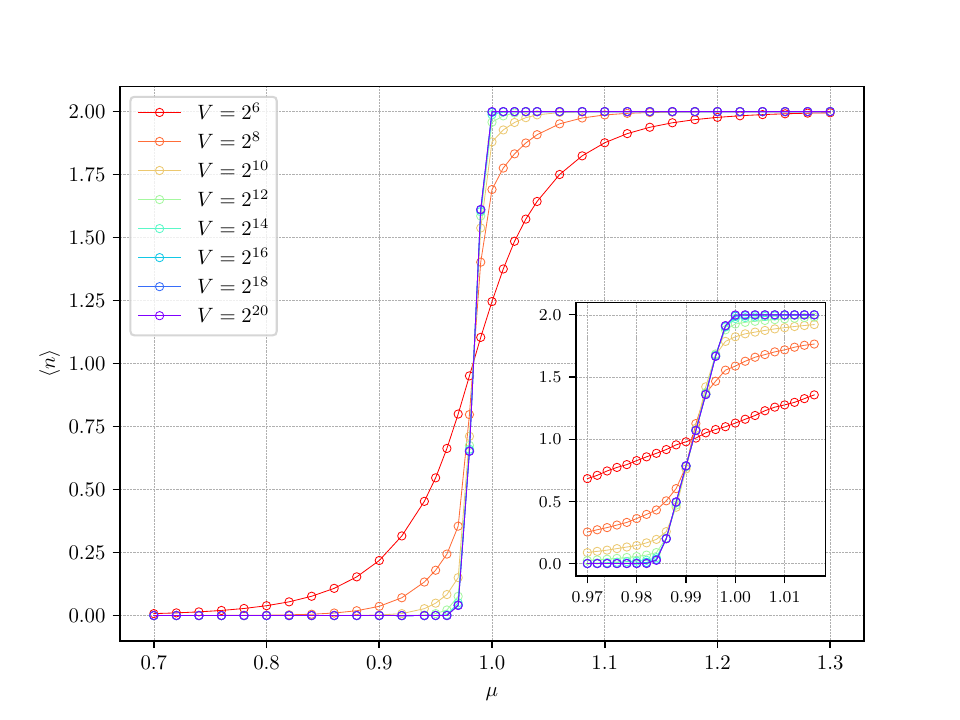}%
            \label{fig:number_m1_beta0_Vdependence}%
        }\hfill
        \subfloat[$\langle \chi \chi \rangle$ as a function of $\mu$ at $m=1$.]{%
            \includegraphics[width=.5\linewidth]{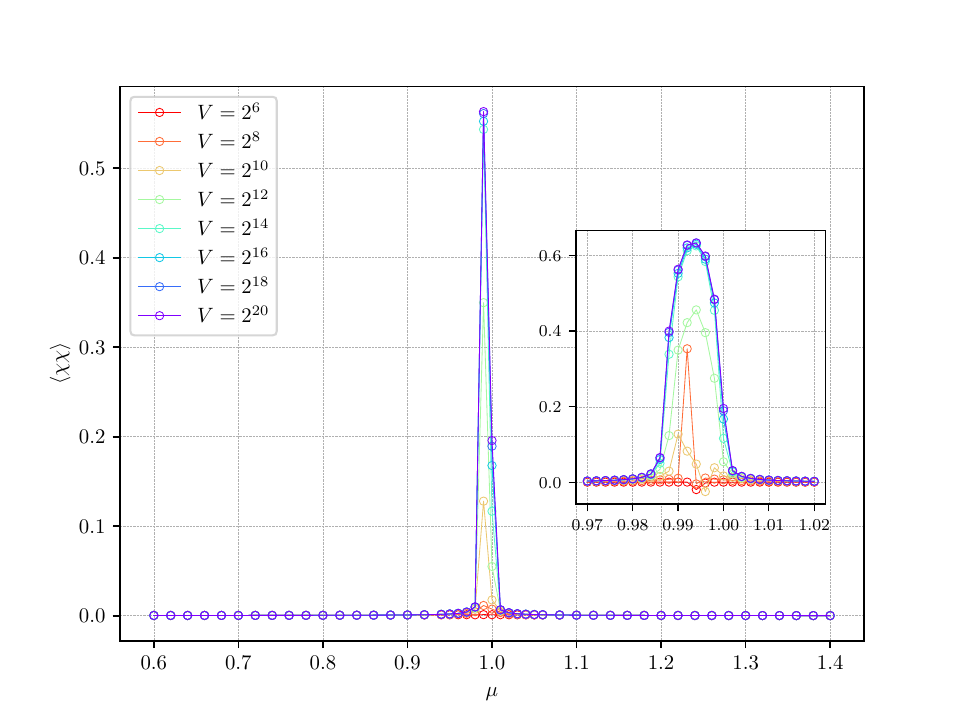}%
            \label{fig:diq_m1_beta0_Vdependence}%
        }
        \caption{Volume dependence of physical quantities in the infinite coupling limit $\beta=0$. The bond dimension in the calculations is $D=84$. To evaluate the numerical differences in Eqs.~\eqref{eq: finite_diff_number} and \eqref{eq: finite_diquark}, we set $\Delta \mu=0.04$ for $m=0.1$, $\Delta \mu=0.02$ for $m=1$, and $\lambda=\Delta\lambda=10^{-4}$. At $m=1$, the insets show the volume dependence of $\langle n \rangle$ and $\langle \chi \chi \rangle$ in the intermediate phase, where $\langle n \rangle$ is evaluated by $\Delta \mu=0.004$.}
        \label{fig:Vdependence_infcoupl}
\end{figure}

We also show the number density $\langle n \rangle$ and diquark condensate $\langle \chi \chi \rangle$ as a function of $\mu$, for $m=0.1$ and $m=1$, at different lattice volumes in Figure \ref{fig:Vdependence_infcoupl}.
For both quark masses, the thermodynamic limit is reached when $V=2^{20}$. 
In the intermediate phase, $\langle \chi \chi \rangle$ increases with the lattice volume until the thermodynamic limit is achieved. As shown in the inset of Figure \ref{fig:diq_m1_beta0_Vdependence}, the negative value of $\langle \chi \chi \rangle$ observed in the intermediate phase might indicate that the current choice of bond dimension ($D=84$) is not large enough for the calculations in small lattice volume. 

\subsection{Finite-$\beta$ regime}

For $\beta>0$, the initial tensor compression scheme is applied before the BTRG calculation. Therefore, we first demonstrate the efficiency of our compression scheme in this section. In Table \ref{table: compress_efficiency_beta1.6}, we can see how the four compressed bond dimensions $D'$ vary with the ratio parameter $r$ defined in Eq.~(\ref{eq:ratio}) at $m=0.1$, $\beta=1.6$, $\mu=0.4$, $\lambda=0$, and $K=14$ as a representative. 
When $r=1$, no compression is made, and we have an original bond dimension that is equal to $16K=224$ with $K=14$.
For an $r$ close enough to one, e.g., $r=0.9999$, the bond dimension of the initial tensors is reduced from $224$ to less than half of its original value after compression. 
The number of tensor elements is only $5.3\%$ of the original one in this case.
We also present another example in Table \ref{table: compress_efficiency_beta0.8}, where $\beta$ is changed to $0.8$. It confirms that our initial tensor compression scheme is more efficient at a smaller $\beta$.
In the following, we always set $r=0.9999$.

\begin{table}
    \centering
    \begin{tabular}{ |c||cccc|c| }
         \hline
         %\multicolumn{4}{|c|}{$m=0.1$, $\beta=1.6$, $\mu=0.4$, $\lambda=0$, $K=14$} \\
         %\hline
         $r$ & $D_1^{'}$ & $D_2^{'}$ & $D_3^{'}$ & $D_4^{'}$ & compression rate \\
         \hline
         1 & $224$ & $224$ & $224$ & $224$ & $100\%$ \\
         0.99999 & $148$ & $148$ & $143$ & $143$ & $17.8\%$ \\
         0.99995 & $122$ & $122$ & $118$ & $118$ & $8.23\%$ \\
         0.9999 & $110$ & $110$ & $105$ & $105$ & $5.30\%$ \\
         0.9995 & $80$ & $80$ & $79$ & $79$ & $1.59\%$ \\
         0.999 & $70$ & $70$ & $67$ & $67$ & $0.874\%$ \\
         0.99 & $35$ & $35$ & $33$ & $33$ & $0.0530\%$ \\
         \hline
    \end{tabular}
    \caption{Efficiency of our initial tensor compression scheme. 
    We set $m=0.1$, $\beta=1.6$, $\mu=0.4$, $\lambda=0$ and $K=14$ as a representative. 
    As shown in Figure~\ref{fig:T_squeezed}, $D_1^{'}$ and $D_2^{'}$ denote the spatial bond dimensions and $D_3^{'}$ and $D_4^{'}$ denote the temporal ones.
    The compression rate in the last column is measured as the number of elements in $\mathcal{T}^{'}_n$ divided by the number of elements in $\mathcal{T}_n$.
    }
    \label{table: compress_efficiency_beta1.6}
\end{table}

\begin{table}
    \centering
    \begin{tabular}{ |c||cccc|c| }
         \hline
         %\multicolumn{4}{|c|}{$m=0.1$, $\beta=1.6$, $\mu=0.4$, $\lambda=0$, $K=14$} \\
         %\hline
         $r$ & $D_1^{'}$ & $D_2^{'}$ & $D_3^{'}$ & $D_4^{'}$ & compression rate \\
         \hline
         1 & $224$ & $224$ & $224$ & $224$ & $100\%$ \\
         0.99999 & $86$ & $86$ & $84$ & $84$ & $2.07\%$ \\
         0.99995 & $68$ & $68$ & $66$ & $66$ & $0.800\%$ \\
         0.9999 & $61$ & $61$ & $59$ & $59$ & $0.514\%$ \\
         0.9995 & $46$ & $46$ & $43$ & $43$ & $0.155\%$ \\
         0.999 & $39$ & $39$ & $37$ & $37$ & $0.0827\%$ \\
         0.99 & $19$ & $19$ & $19$ & $19$ & $0.00518\%$ \\
         \hline
    \end{tabular}
    \caption{Another example illustrating the efficiency of our initial tensor compression scheme with $m=0.1$, $\beta=0.8$, $\mu=0.4$, $\lambda=0$ and $K=14$.
    As shown in Figure~\ref{fig:T_squeezed}, $D_1^{'}$ and $D_2^{'}$ denote the spatial bond dimensions and $D_3^{'}$ and $D_4^{'}$ denote the temporal ones.
    The compression rate in the last column is measured as the number of elements in $\mathcal{T}^{'}_n$ divided by the number of elements in $\mathcal{T}_n$.
    }
    \label{table: compress_efficiency_beta0.8}
\end{table}

\begin{figure}[htbp]
    \centering
    \includegraphics[width=0.7\textwidth]{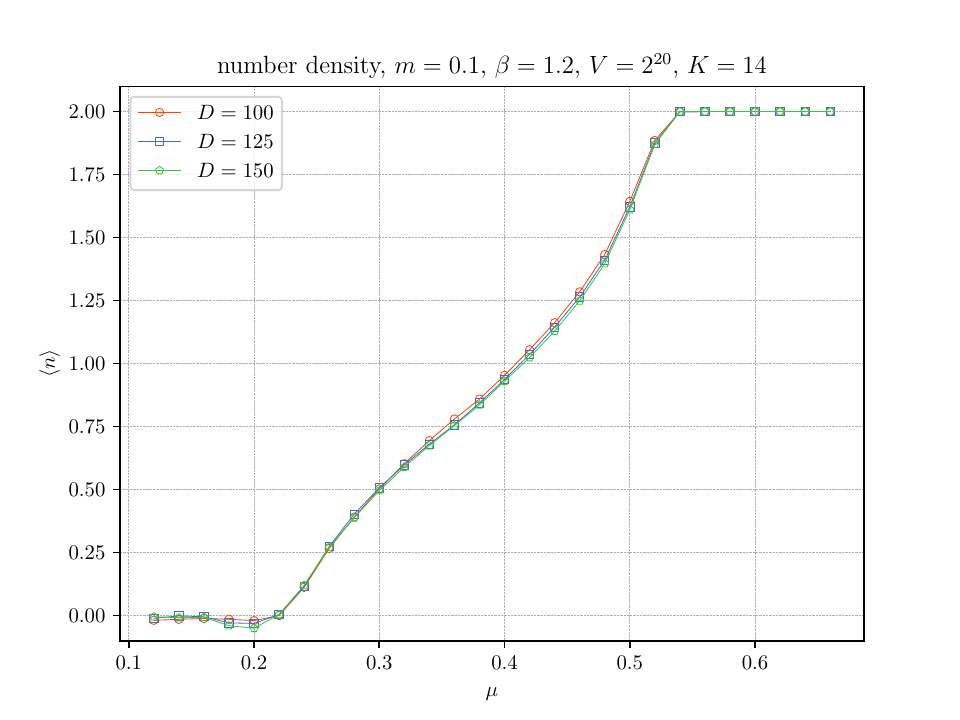}
    \caption{Quark number density $\langle n \rangle$ as a function of chemical potential $\mu$ at $m=0.1$, $\beta=1.2$, in the thermodynamic limit. We set the bond dimensions $D=100, 125$, and $150$. The sample size for the discretization of gauge group integrations is $K=14$. To evaluate the numerical difference in Eq.~\eqref{eq: finite_diff_number}, we set $\Delta \mu=0.04$.}
    \label{fig:m0.1_beta1.2_D}
\end{figure}

Next, we check the algorithmic parameter dependence of the BTRG results. 
At $m=0.1$, $\beta=1.2$ and $V=2^{20}$, we calculate $\langle n \rangle$ as a function of $\mu$ using different bond dimension $D$ as shown in Figure \ref{fig:m0.1_beta1.2_D}, with a fixed number of sampled $SU(2)$ matrices $K=14$. The qualitative behavior of $\langle n \rangle$ and the position of the two transition points $\mu_{c1/c2}$ are showing consistency as $D \geq 100$, despite some small deviations in the numerical values of $\langle n \rangle$.

\begin{figure}
    \centering
    \begin{subfigure}[h]{0.7\textwidth}
        \captionsetup{justification=centering}
        \centering
        \includegraphics[width=\linewidth]{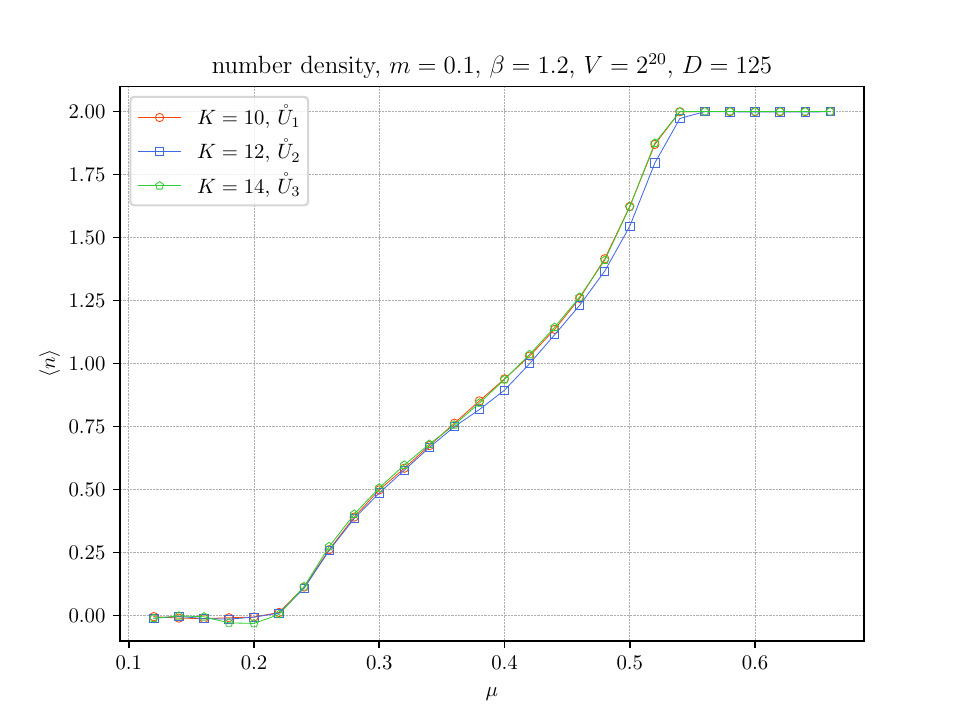} 
        \caption{} \label{fig:m0.1_beta1.2_K}
    \end{subfigure} \\
    \begin{subfigure}[h]{0.7\textwidth}
        \captionsetup{justification=centering}
        \centering
        \includegraphics[width=\linewidth]{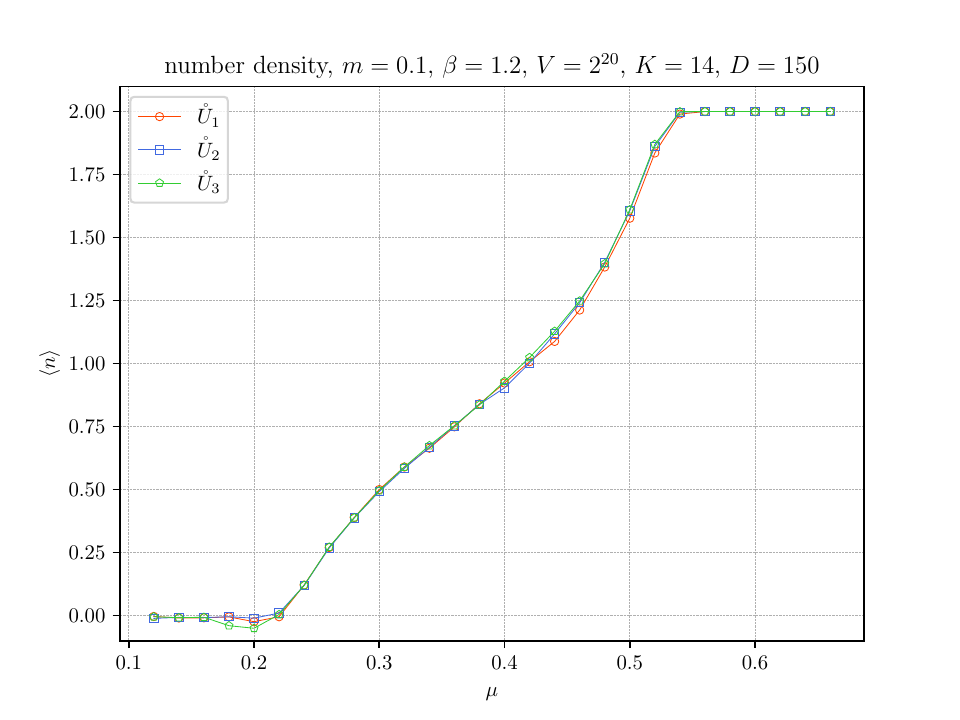} 
        \caption{} \label{fig:m0.1_beta1.2_gseeds}
    \end{subfigure}
    \caption{Quark number density $\langle n \rangle$ as a function of chemical potential $\mu$ at $m=0.1$, $\beta=1.2$, in the thermodynamic limit. We set $\Delta\mu=0.04$ to evaluate the numerical difference in Eq.~\eqref{eq: finite_diff_number}. (\protect\subref{fig:m0.1_beta1.2_K}): The sample sizes for the discretization of gauge group integrations are set to be $K=10$, $12$, and $14$, with a distinct random matrix set $\mathring{U}_i$ chosen for each $K$. The bond dimension is set to be $D=125$. (\protect\subref{fig:m0.1_beta1.2_gseeds}): The sample size is fixed to be $K=14$, and the calculation is repeated using the same three matrix sets. The bond dimension is set to be $D=150$.
    \label{fig:m0.1_beta1.2_parameters}}
\end{figure}

Similarly, $\langle n \rangle$ at $m=0.1$, $\beta=1.2$ and $V=2^{20}$ calculated with different $K$ and various matrix sets $\mathring{U}_i$ are shown in Figure \ref{fig:m0.1_beta1.2_K} with $D=125$. 
The results suggest that $K=14$ suffices for our purpose and $\langle n \rangle$ obtained from different $\mathring{U}_i$ exhibit similar qualitative behavior. Therefore, we always set $K=14$ and $D=150$ for the finite-$\beta$ calculations in this study.

$\langle n \rangle$ at the same $m$, $\beta$ and $V$, calculated with a fixed $K=14$ and $D=150$, using the three matrix sets adopted in Figure \ref{fig:m0.1_beta1.2_K}, are illustrated in Figure \ref{fig:m0.1_beta1.2_gseeds}. 
Comparing Figure \ref{fig:m0.1_beta1.2_K} and \ref{fig:m0.1_beta1.2_gseeds}, the results near $\mu_{c2}$ are affected more by the sample size $K$, rather than the choice of random matrix set. 
On the other hand, the results in the Silver-Blaze region rely more on which matrix set is being picked. 
Negative $\langle n \rangle$ is seen around $\mu=0.2$, presumably because a finite number of $SU(2)$ matrices are used to approximate the gauge group integration. 
Since the $SU(2)$ matrices in each matrix set are randomly sampled, it is normal that some particular matrix set is able to give a better approximation under a fixed and finite sample size.
For example, we can see from Figure \ref{fig:m0.1_beta1.2_parameters} that the negative $\langle n \rangle$ issue is eased when the matrix set $\mathring{U}_2$ is picked.

We also comment on the practice of taking a statistical average over trials using distinct sets of $SU(2)$ matrices. 
It is believed to be able to reduce the systematic bias due to the finite-$K$ effect and improve the accuracy of the data such as reproducing the Silver-Blaze phenomenon better in the small $\mu$ regime. However, it is computationally demanding; we have found that so many trials are required to stabilize the numerical results, particularly the numerical differences. 
Therefore, we employ the same matrix set at different $\mu$ for a specific observable at a particular $(m, \beta)$.

\begin{figure}[htbp]
    \centering
    \includegraphics[scale=0.8]{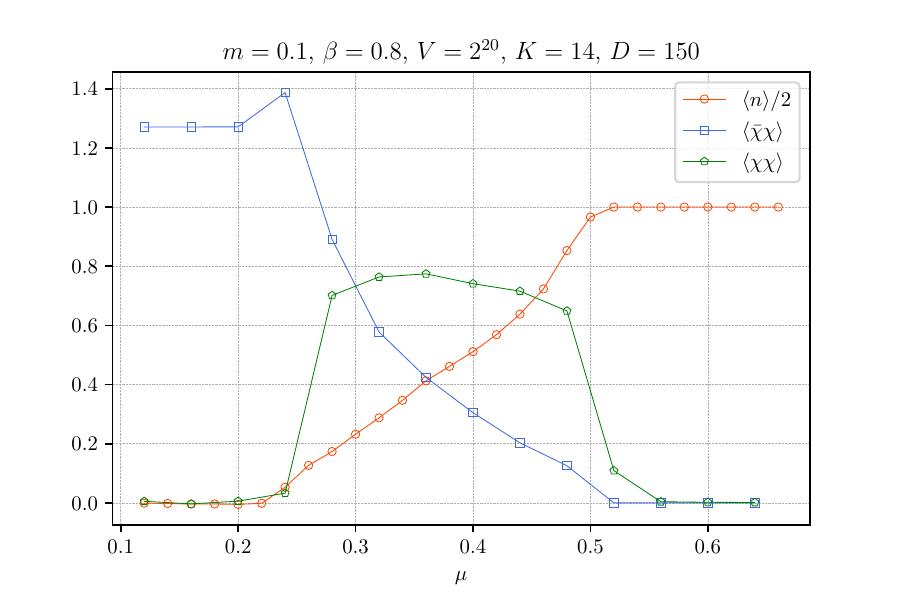}
    \caption{Quark number density $\langle n \rangle$, fermion condensate $\langle \bar{\chi} \chi \rangle$ and diquark condensate $\langle \chi \chi \rangle$ as a function of chemical potential $\mu$ at $m=0.1$, $\beta=0.8$, in the thermodynamic limit. The bond dimension in the calculations is $D=150$. The sample size for the discretization of gauge group integrations is $K=14$. To evaluate the numerical differences in Eqs.~\eqref{eq: finite_diff_number}, \eqref{eq: finite_diff_fermion}, and \eqref{eq: finite_diquark}, we set $\Delta \mu=0.04$, $\Delta m = 10^{-4}$, and $\lambda=\Delta\lambda=10^{-4}$.}
    \label{fig:m0.1_beta0.8}
\end{figure}

\begin{figure}[htbp]
    \centering
    \includegraphics[scale=0.8]{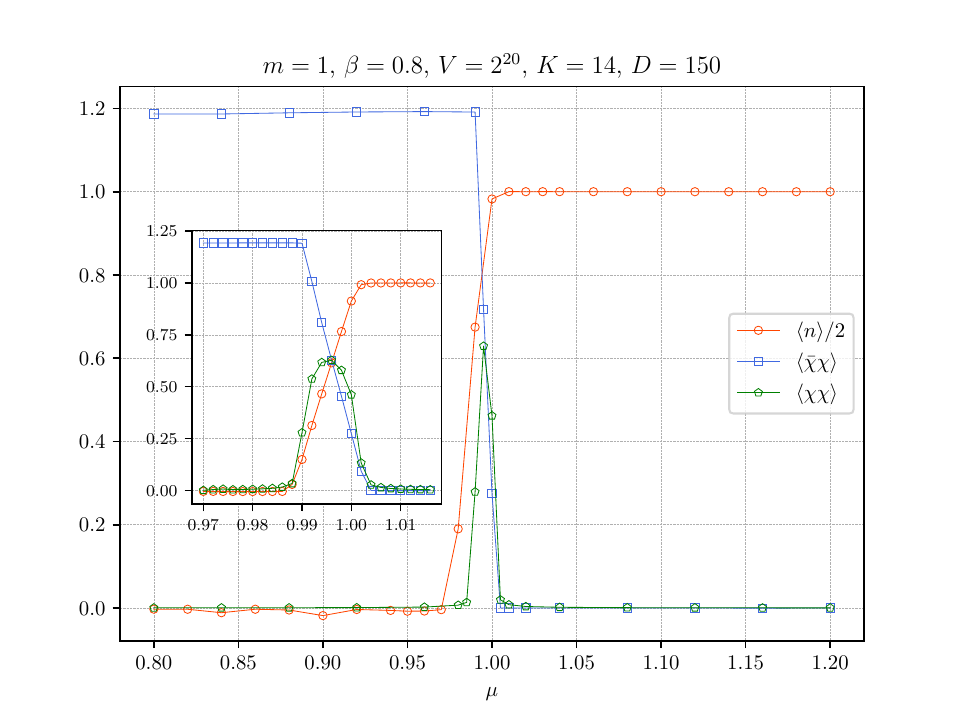}
    \caption{Quark number density $\langle n \rangle$, fermion condensate $\langle \bar{\chi} \chi \rangle$ and diquark condensate $\langle \chi \chi \rangle$ as a function of chemical potential $\mu$ at $m=1$, $\beta=0.8$, in the thermodynamic limit. The bond dimension in the calculations is $D=150$. The sample size for the discretization of gauge group integrations is $K=14$. To evaluate the numerical differences in Eqs.~\eqref{eq: finite_diff_number}, \eqref{eq: finite_diff_fermion}, and \eqref{eq: finite_diquark}, we set $\Delta \mu=0.02$, $\Delta m = 10^{-4}$, and $\lambda=\Delta\lambda=10^{-4}$. The inset shows the three quantities in the intermediate phase, where $\langle n \rangle$ is evaluated using a finer $\Delta \mu = 0.004$.}
    \label{fig:m1_beta0.8}
\end{figure}

\begin{figure}[ht]
    \captionsetup[subfigure]{justification=centering}
        \subfloat[$\langle n \rangle$ as a function of $\mu$ at $m=0.1$.]{%
            \includegraphics[width=.5\linewidth]{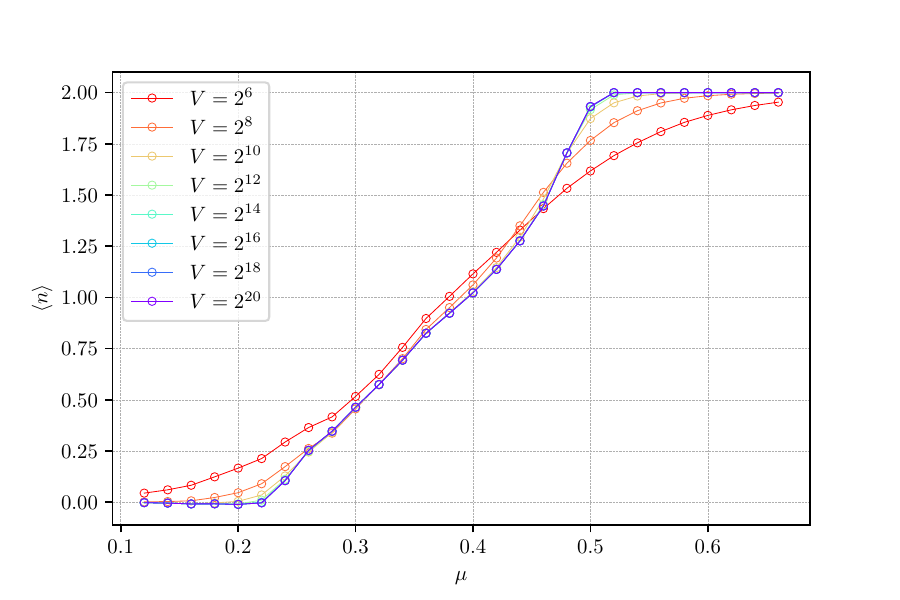}%
            \label{fig:number_m0.1_beta0.8_Vdependence}%
        }\hfill
        \subfloat[$\langle \chi \chi \rangle$ as a function of $\mu$ at $m=0.1$.]{%
            \includegraphics[width=.5\linewidth]{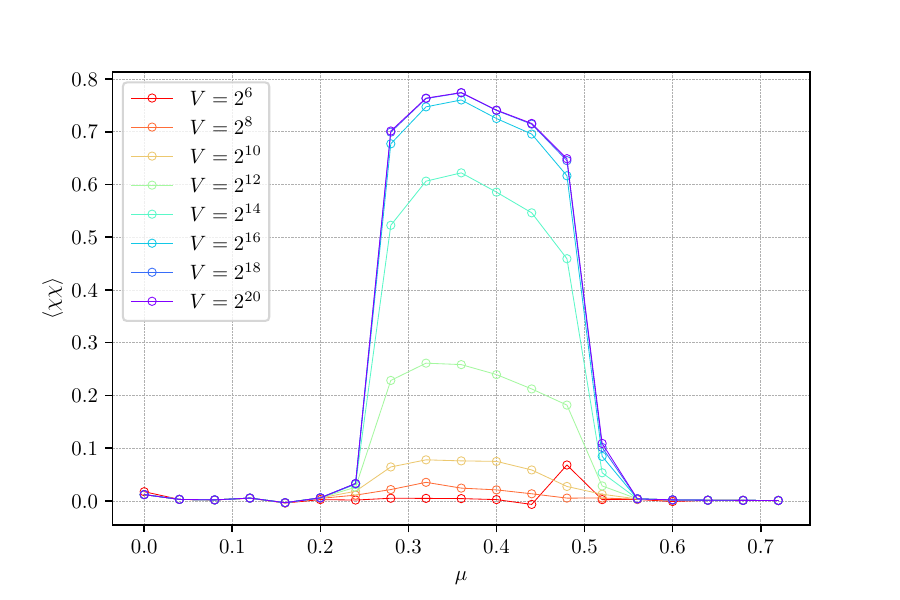}%
            \label{fig:diq_m0.1_beta0.8_Vdependence}%
        }\\
        \subfloat[$\langle n \rangle$ as a function of $\mu$ at $m=1$.]{%
            \includegraphics[width=.5\linewidth]{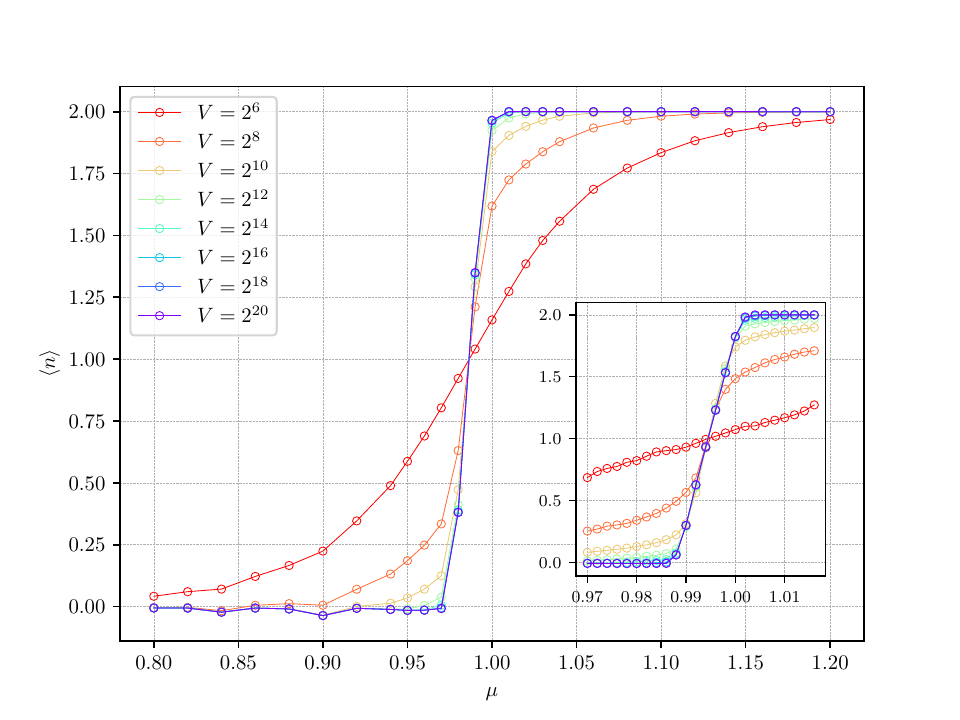}%
            \label{fig:number_m1_beta0.8_Vdependence}%
        }\hfill
        \subfloat[$\langle \chi \chi \rangle$ as a function of $\mu$ at $m=1$.]{%
            \includegraphics[width=.5\linewidth]{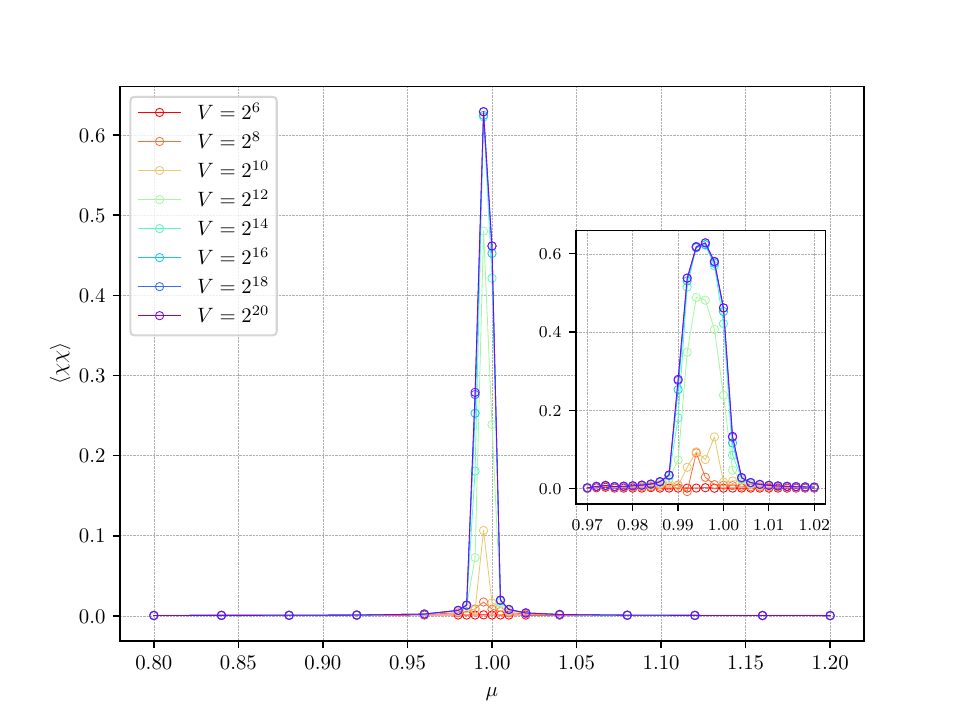}%
            \label{fig:diq_m1_beta0.8_Vdependence}%
        }
        \caption{Volume dependence of physical quantities at $\beta=0.8$. The bond dimension in the calculations is $D=150$. The sample size for the discretization of gauge group integrations is $K=14$. To evaluate the numerical differences in Eqs.~\eqref{eq: finite_diff_number} and \eqref{eq: finite_diquark}, we set $\Delta \mu=0.04$ for $m=0.1$, $\Delta \mu=0.02$ for $m=1$, and $\lambda=\Delta\lambda=10^{-4}$. At $m=1$, the insets show the volume dependence of $\langle n \rangle$ and $\langle \chi \chi \rangle$ in the intermediate phase, where $\langle n \rangle$ is evaluated by $\Delta\mu=0.004$.}
        \label{fig:Vdependence_beta0.8}
\end{figure}

Now, we see the numerical results of $\langle n \rangle$, $\langle \bar{\chi} \chi \rangle$ and $\langle \chi \chi \rangle$ as a function of $\mu$ in the thermodynamic limit $V=2^{20}$, at a finite $\beta=0.8$. For $m=0.1$, the behavior of the three observables shown in Figure \ref{fig:m0.1_beta0.8} is similar to that at $\beta=0$. We observe a more extended intermediate phase in $\mu$ as $\beta$ becomes non-zero. At $m=0.1$ and $\beta=0.8$, $\mu_{c1}=0.22$ and $\mu_{c2}=0.52$. For $m=1$, we see a sharp transition in Figure \ref{fig:m1_beta0.8} as in the infinite coupling limit while the intermediate phase shifts slightly in $\mu$ with $\mu_{c1} \approx 0.986$ and $\mu_{c2} \approx 1.004$. From the inset of Figure \ref{fig:m1_beta0.8}, the crossover nature persists even when $\beta$ becomes finite. The volume dependence of $\langle n \rangle$ and $\langle \chi \chi \rangle$ at $m=0.1$ and $m=1$ are shown in Figure \ref{fig:Vdependence_beta0.8}.

\begin{figure}[htbp]
    \centering
    \includegraphics[scale=0.8]{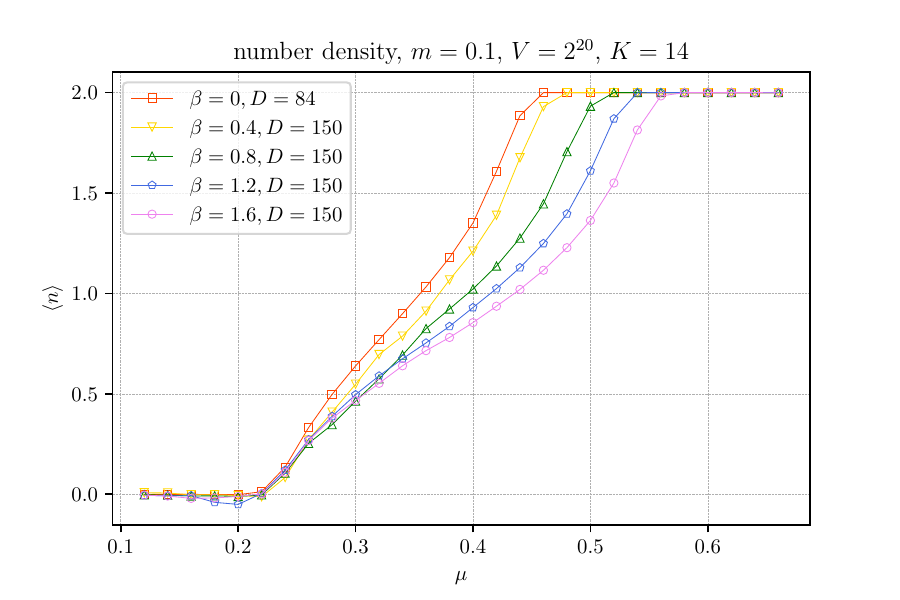}
    \caption{Quark number density $\langle n \rangle$ as a function of chemical potential $\mu$ at $m=0.1$, $\beta=0, 0.4, 0.8, 1.2, 1.6$ in the thermodynamic limit. The bond dimension in the infinite coupling calculation is $D=84$, and the bond dimension in the finite $\beta$ calculations is $D=150$. At a finite $\beta$, the sample size for the discretization of gauge group integrations is $K=14$. To evaluate the numerical differences in Eq.~\eqref{eq: finite_diff_number}, we set $\Delta \mu=0.04$.}
    \label{fig:number_betas}
\end{figure}

Finally, we describe the $\beta$ dependence for $\mu_{c1}$ and $\mu_{c2}$ at $m=0.1$. As shown in Figure \ref{fig:number_betas}, the first transition point appears to be robust against $\beta$, i.e., $\mu_{c1} \approx 0.22$ for $0 \le \beta \le 1.6$ at $m=0.1$. We expect that $\mu_{c1}$ becomes smaller as $\beta$ increases because the mass gap vanishes in the limit of $\beta\to\infty$.
To observe such behavior, we might need to compute the number density with $\beta>1.6$ or to employ finer $\Delta\mu$ to improve the accuracy of the finite difference in Eq.~\eqref{eq: finite_diff_number}.
On the other hand, the second transition point $\mu_{c2}$ is located at larger $\mu$ as $\beta$ increases. It is expected because $\langle n \rangle$ does not saturate in regions of larger $\mu$ as the gauge coupling is weakened, namely approaching the continuum limit.

\section{Conclusion}

In this work, we have investigated the $(1+1)$-dimensional two-color QCD at finite density with staggered fermions using the TRG approach.
We have used a random sampling method to discretize the gauge group integration and construct a Grassmann tensor network representation for the partition function. 
Since TRG calculations for non-Abelian gauge theories coupled with fermions are computationally challenging due to the very large initial bond dimension, we have proposed an efficient initial tensor compression scheme that can also be applied to other lattice models. 
We have evaluated the expectation values of the quark number density, fermion condensate, and diquark condensate.
These observables are widely employed to investigate the phase structure of two-color QCD in higher dimensions.
We have made BTRG calculations both at the infinite coupling limit and the finite coupling regime. 
Since there is no spontaneous breaking of continuous global symmetry in two dimensions, the fermion condensate and diquark condensate are computed by explicitly breaking the $U(1)_A$ and $U(1)_V$. 
Under this setting, we have found that the behavior of the observables calculated by the TRG approach is similar to that reported in a previous mean-field study~\cite{Nishida:2003uj} for the higher-dimensional two-color QCD. 
We have also studied how the positions of the two transition points vary as the gauge coupling changes.

Our results encourage a future application of the TRG approach to the higher-dimensional two-color QCD. 
It is possible to improve our construction of initial tensors such that more $SU(2)$ matrices are used to discretize the gauge group integration, i.e., a larger $K$ is allowed, without increasing the memory requirement significantly. In higher-dimensional cases where spontaneous breaking of continuous symmetry can exist, it is instructive to apply the TRG approach to evaluate the order parameters and perform extrapolations toward the chiral limit ($m=0$) and the vanishing $\lambda$ limit.
We also emphasize that our Grassmann tensor network representation for the partition function can be extended to the three-color theory, which suffers from the sign problem at finite density, without conceptual difficulties.

Finally, we note that inhomogeneous phases in two-color QCD and related models~\cite{Schon:2000he, Thies:2003kk, Kojo:2011fh, Hayata:2023pkw} have attracted attention recently. 
As another future direction, we will explore the applicability of the TRG approach in studying the spatial dependence of physical quantities.

\FloatBarrier

\begin{acknowledgments}
A part of the numerical calculation for the present work was carried out with ohtaka provided by the Institute for Solid State Physics, the University of Tokyo. 
This work is supported by the Endowed Project for Quantum Software Research and Education, the University of Tokyo~\cite{qsw}, and the Center of Innovations for Sustainable Quantum AI (JST Grant Number JPMJPF2221).
SA acknowledges the support from JSPS KAKENHI (JP23K13096, JP24H00214) and the Top Runners in Strategy of Transborder Advanced Researches (TRiSTAR) program conducted as the Strategic Professional Development Program for Young Researchers by the MEXT.
\end{acknowledgments}

\appendix

\section{Coefficients of the Grassmann tensor $\mathcal{F}$}
\label{app:elts_F}
Here, we discuss one method to derive the tensor elements of $F$ in Eq.~\eqref{eq:Tf2}. 
For simplicity, we consider the two-color case where $N=2$. 
We also label the spacetime direction $\nu$ by $\nu=x,t$ not by $\nu=1,2$.
In this case, the integration over the original staggered fermions is written as
\begin{align} 
\label{eq:elmt_tf}
    \mathcal{F} 
    &= \int {\rm d}\chi_1 {\rm d}\bar{\chi}_1 {\rm d}\chi_2 {\rm d}\bar{\chi}_2 \, 
    \text{e}^{-m \left(\bar{\chi}_1 \chi_1 + \bar{\chi}_2 \chi_2\right)} \left(1+\bar{\chi}_1 A \right)  \left(1+\chi_1 B \right) \left(1+\bar{\chi}_2 C \right)  \left(1+\chi_2 D \right) \nonumber\\
    &= ABCD + mCD +mAB + m^2 ,
\end{align}
where $A$, $B$, $C$, and $D$ are sums of terms with one auxiliary Grassmann variable, and their expressions are given by
\begin{align}
    A &= -\eta_{x,1} -\eta_{t,1} + (U_x^\dagger)_{11} \bar{\zeta}_{x,1} + (U_x^\dagger)_{12} \bar{\zeta}_{x,2} + (U_t^\dagger)_{11} \bar{\zeta}_{t,1} + (U_t^\dagger)_{12} \bar{\zeta}_{t,2} ,
\end{align}
\begin{align}
    B &= \frac{1}{2} \Big[ 
        - (U_x)_{11} \bar{\eta}_{x,1}  
        - (U_x)_{21} \bar{\eta}_{x,2} 
        - a_{+} (U_t)_{11} \bar{\eta}_{t,1} 
        %- \text{e}^{\mu} p_{t}(n) (U_t)_{11} \bar{\eta}_{t,1} 
        - a_{+} (U_t)_{21} \bar{\eta}_{t,2} 
        %- \text{e}^{\mu} p_{t}(n) (U_t)_{21} \bar{\eta}_{t,2} 
        + \zeta_{x,1}   
        + a_{-} \zeta_{t,1} \Big] ,
        %+ \text{e}^{-\mu} p_{t}(n) \zeta_{t,1} \Big] ,
\end{align}
\begin{align}
    C &= -\eta_{x,2} -\eta_{t,2} + (U_x^\dagger)_{21} \bar{\zeta}_{x,1} + (U_x^\dagger)_{22} \bar{\zeta}_{x,2} + (U_t^\dagger)_{21} \bar{\zeta}_{t,1} + (U_t^\dagger)_{22} \bar{\zeta}_{t,2} ,
\end{align}
\begin{align}
    D &= \frac{1}{2} \Big[ 
    -(U_x)_{12} \bar{\eta}_{x,1}  
    - (U_x)_{22} \bar{\eta}_{x,2} 
    - a_{+} (U_t)_{12} \bar{\eta}_{t,1} 
    %- \text{e}^{\mu} (-1)^{n_x} (U_t)_{12} \bar{\eta}_{t,1} 
    - a_{+} (U_t)_{22} \bar{\eta}_{t,2} 
    %- \text{e}^{\mu} (-1)^{n_x} (U_t)_{22} \bar{\eta}_{t,2} 
    + \zeta_{x,2}   
    + a_{-} \zeta_{t,2} \Big] ,
    %+ \text{e}^{-\mu} (-1)^{n_x} \zeta_{t,2} \Big] ,
\end{align}
with $a_{\pm}={\rm e}^{\pm\mu}p_{t}(n)$.
After expanding out the second line in Eq.~\eqref{eq:elmt_tf}, one can compare like terms in Eqs.~\eqref{eq:elmt_tf} and \eqref{eq:Tf2}, then obtain the elements of coefficient tensor $F$.
When a diquark source term $\sum_{n} \lambda \left[\chi^T(n) \sigma_2 \chi(n) + \bar{\chi}(n) \sigma_2 \bar{\chi}^T(n) \right]/2$ is added to the action, then the Grassmann tensor $\mathcal{F}$ in Eq.~\eqref{eq:Tf2} is given by $ABCD + m(AB+CD) + {\rm i} \lambda (AC + BD) + m^2 + \lambda^2$.

\section{Construction of $\rho$}
\label{app: make_rho}

In section 3.2, we introduced a Hermitian matrix as $\rho_B = M_B^\dagger M_B$. It is computationally demanding to obtain $\rho_B$ directly through this formula since it is a multiplication between a $16K \times 16^3K^3$ matrix and a $16^3K^3 \times 16K$ matrix.

In this study, we obtain $\rho_B$ (and $\rho_A$ by similar steps) following another equivalent method. First, we perform the following SVD on the coefficient tensor $T_n$ as
\begin{equation} 
\label{eq: B1}
(-1)^{f_{x}(f_{t}+f_{x'})+f_{x'}f_t} (T_n)_{x t x' t'} = \sum_{y=1}^{16^2 K^2} (U_B)_{x'ty} (s_B)_{y} (V_B^\dagger)_{yxt'}.
\end{equation}
Since the direct calculation of Eq.~(\ref{eq: B1}) is also demanding, we employ a randomized SVD~\cite{PhysRevE.97.033310} with the oversampling parameter $p \sim 0.07 (16K)^2$ and the iteration number of QR decomposition $r'=7$ in this step.

Then we substitute Eq.~(\ref{eq: B1}) into the definition of $\rho_B$:
\begin{equation} 
\label{eq: B2}
\begin{split}
(\rho_B)_{x\tilde{x}} &= \sum_{t,x',t'} (M^*_B)_{(t x' t') x} (M_B)_{(t x' t') \tilde{x}} \\
&= \sum_{t,x',t'} (-1)^{f_{x}(f_{t}+f_{x'})+f_{x'}f_t} (T^*_n)_{x t x' t'} (-1)^{f_{\tilde{x}}(f_{t}+f_{x'})+f_{x'}f_t} (T_n)_{\tilde{x} t x' t'} \\
&= \sum_{t,x',t'} \sum_{y,\tilde{y}} (U_B)^*_{x'ty} (s_B)_{y} (V_B^\dagger)^*_{yxt'} (U_B)_{x't\tilde{y}} (s_B)_{\tilde{y}} (V_B^\dagger)_{\tilde{y}\tilde{x}t'}  \\
&= \sum_{t'} \sum_{y,\tilde{y}} \delta_{y,\tilde{y}} (s_B)_{y} (V_B^\dagger)^*_{yxt'} (s_B)_{\tilde{y}} (V_B^\dagger)_{\tilde{y}\tilde{x}t'} \\
&= \sum_{t'} \sum_{y} (s_B)^2_{y} (V_B^\dagger)^*_{yxt'} (V_B^\dagger)_{y\tilde{x}t'}.
\end{split}
\end{equation}
From the third to the fourth line of Eq.~(\ref{eq: B2}), we used the property $U_B^\dagger U_B=I$ of an SVD. Eq.~(\ref{eq: B2}) shows that $\rho_B$ can be made once $s_B$ and $V_B^\dagger$ in Eq.~(\ref{eq: B1}) are obtained.

\bibliographystyle{JHEP.bst}
\bibliography{bib/for_this_paper,bib/formulation,bib/grassmann,bib/gauge,bib/algorithm,bib/continuous}

\providecommand{\href}[2]{#2}\begingroup\raggedright\begin{thebibliography}{10}

\bibitem{Levin:2006jai}
M.~Levin and C.~P. Nave, \emph{{Tensor renormalization group approach to two-dimensional classical lattice models}}, \href{https://doi.org/10.1103/PhysRevLett.99.120601}{\emph{Phys. Rev. Lett.} {\bfseries 99} (2007) 120601}, [\href{https://arxiv.org/abs/cond-mat/0611687}{{\ttfamily cond-mat/0611687}}].

\bibitem{Xie:2009zzd}
Z.~Y. Xie, H.~C. Jiang, Q.~N. Chen, Z.~Y. Weng and T.~Xiang, \emph{{Second Renormalization of Tensor-Network States}}, \href{https://doi.org/10.1103/PhysRevLett.103.160601}{\emph{Phys. Rev. Lett.} {\bfseries 103} (2009) 160601}, [\href{https://arxiv.org/abs/0809.0182}{{\ttfamily 0809.0182}}].

\bibitem{PhysRevLett.115.180405}
G.~Evenbly and G.~Vidal, \emph{Tensor network renormalization}, \href{https://doi.org/10.1103/PhysRevLett.115.180405}{\emph{Phys. Rev. Lett.} {\bfseries 115} (Oct, 2015) 180405}.

\bibitem{PhysRevLett.118.110504}
S.~Yang, Z.-C. Gu and X.-G. Wen, \emph{Loop optimization for tensor network renormalization}, \href{https://doi.org/10.1103/PhysRevLett.118.110504}{\emph{Phys. Rev. Lett.} {\bfseries 118} (Mar, 2017) 110504}.

\bibitem{PhysRevE.97.033310}
S.~Morita, R.~Igarashi, H.-H. Zhao and N.~Kawashima, \emph{Tensor renormalization group with randomized singular value decomposition}, \href{https://doi.org/10.1103/PhysRevE.97.033310}{\emph{Phys. Rev. E} {\bfseries 97} (Mar, 2018) 033310}.

\bibitem{PhysRevB.105.L060402}
D.~Adachi, T.~Okubo and S.~Todo, \emph{Bond-weighted tensor renormalization group}, \href{https://doi.org/10.1103/PhysRevB.105.L060402}{\emph{Phys. Rev. B} {\bfseries 105} (Feb, 2022) L060402}, [\href{https://arxiv.org/abs/2011.01679}{{\ttfamily 2011.01679}}].

\bibitem{PhysRevB.86.045139}
Z.~Y. Xie, J.~Chen, M.~P. Qin, J.~W. Zhu, L.~P. Yang and T.~Xiang, \emph{Coarse-graining renormalization by higher-order singular value decomposition}, \href{https://doi.org/10.1103/PhysRevB.86.045139}{\emph{Phys. Rev. B} {\bfseries 86} (Jul, 2012) 045139}, [\href{https://arxiv.org/abs/1201.1144}{{\ttfamily 1201.1144}}].

\bibitem{Adachi:2019paf}
D.~Adachi, T.~Okubo and S.~Todo, \emph{{Anisotropic Tensor Renormalization Group}}, \href{https://doi.org/10.1103/PhysRevB.102.054432}{\emph{Phys. Rev. B} {\bfseries 102} (2020) 054432}, [\href{https://arxiv.org/abs/1906.02007}{{\ttfamily 1906.02007}}].

\bibitem{Kadoh:2019kqk}
D.~Kadoh and K.~Nakayama, \emph{{Renormalization group on a triad network}},  \href{https://arxiv.org/abs/1912.02414}{{\ttfamily 1912.02414}}.

\bibitem{Yamashita:2021yxs}
T.~Yamashita and T.~Sakurai, \emph{{A parallel computing method for the higher order tensor renormalization group}}, \href{https://doi.org/10.1016/j.cpc.2022.108423}{\emph{Comput. Phys. Commun.} {\bfseries 278} (2022) 108423}, [\href{https://arxiv.org/abs/2110.03607}{{\ttfamily 2110.03607}}].

\bibitem{Nakayama:2023ytr}
K.~Nakayama, \emph{{Randomized higher-order tensor renormalization group}},  \href{https://arxiv.org/abs/2307.14191}{{\ttfamily 2307.14191}}.

\bibitem{Gu:2010yh}
Z.-C. Gu, F.~Verstraete and X.-G. Wen, \emph{{Grassmann tensor network states and its renormalization for strongly correlated fermionic and bosonic states}},  \href{https://arxiv.org/abs/1004.2563}{{\ttfamily 1004.2563}}.

\bibitem{Gu:2013gba}
Z.-C. Gu, \emph{{Efficient simulation of Grassmann tensor product states}}, \href{https://doi.org/10.1103/PhysRevB.88.115139}{\emph{Phys. Rev.} {\bfseries B88} (2013) 115139}, [\href{https://arxiv.org/abs/1109.4470}{{\ttfamily 1109.4470}}].

\bibitem{Shimizu:2014uva}
Y.~Shimizu and Y.~Kuramashi, \emph{{Grassmann tensor renormalization group approach to one-flavor lattice Schwinger model}}, \href{https://doi.org/10.1103/PhysRevD.90.014508}{\emph{Phys. Rev.} {\bfseries D90} (2014) 014508}, [\href{https://arxiv.org/abs/1403.0642}{{\ttfamily 1403.0642}}].

\bibitem{Akiyama:2020sfo}
S.~Akiyama and D.~Kadoh, \emph{{More about the Grassmann tensor renormalization group}}, \href{https://doi.org/10.1007/JHEP10(2021)188}{\emph{JHEP} {\bfseries 10} (2021) 188}, [\href{https://arxiv.org/abs/2005.07570}{{\ttfamily 2005.07570}}].

\bibitem{Shimizu:2012zza}
Y.~Shimizu, \emph{{Tensor renormalization group approach to a lattice boson model}}, \href{https://doi.org/10.1142/S0217732312500356}{\emph{Mod. Phys. Lett.} {\bfseries A27} (2012) 1250035}.

\bibitem{Shimizu:2012wfa}
Y.~Shimizu, \emph{{Analysis of the (1+1)-dimensional lattice $\phi^{4}$ model using the tensor renormalization group}}, {\emph{Chin. J. Phys.} {\bfseries 50} (2012) 749}.

\bibitem{Kadoh:2018tis}
D.~Kadoh, Y.~Kuramashi, Y.~Nakamura, R.~Sakai, S.~Takeda and Y.~Yoshimura, \emph{{Tensor network analysis of critical coupling in two dimensional $\phi^{4}$ theory}}, \href{https://doi.org/10.1007/JHEP05(2019)184}{\emph{JHEP} {\bfseries 05} (2019) 184}, [\href{https://arxiv.org/abs/1811.12376}{{\ttfamily 1811.12376}}].

\bibitem{Kadoh:2019ube}
D.~Kadoh, Y.~Kuramashi, Y.~Nakamura, R.~Sakai, S.~Takeda and Y.~Yoshimura, \emph{{Investigation of complex $\phi^{4}$ theory at finite density in two dimensions using TRG}}, \href{https://doi.org/10.1007/JHEP02(2020)161}{\emph{JHEP} {\bfseries 02} (2020) 161}, [\href{https://arxiv.org/abs/1912.13092}{{\ttfamily 1912.13092}}].

\bibitem{Delcamp:2020hzo}
C.~Delcamp and A.~Tilloy, \emph{{Computing the renormalization group flow of two-dimensional $\phi^4$ theory with tensor networks}}, \href{https://doi.org/10.1103/PhysRevResearch.2.033278}{\emph{Phys. Rev. Res.} {\bfseries 2} (2020) 033278}, [\href{https://arxiv.org/abs/2003.12993}{{\ttfamily 2003.12993}}].

\bibitem{Akiyama:2020ntf}
S.~Akiyama, D.~Kadoh, Y.~Kuramashi, T.~Yamashita and Y.~Yoshimura, \emph{{Tensor renormalization group approach to four-dimensional complex $\phi^4$ theory at finite density}}, \href{https://doi.org/10.1007/JHEP09(2020)177}{\emph{JHEP} {\bfseries 09} (2020) 177}, [\href{https://arxiv.org/abs/2005.04645}{{\ttfamily 2005.04645}}].

\bibitem{Akiyama:2021zhf}
S.~Akiyama, Y.~Kuramashi and Y.~Yoshimura, \emph{{Phase transition of four-dimensional lattice $\phi^4$ theory with tensor renormalization group}}, \href{https://doi.org/10.1103/PhysRevD.104.034507}{\emph{Phys. Rev. D} {\bfseries 104} (2021) 034507}, [\href{https://arxiv.org/abs/2101.06953}{{\ttfamily 2101.06953}}].

\bibitem{Shimizu:2014fsa}
Y.~Shimizu and Y.~Kuramashi, \emph{{Critical behavior of the lattice Schwinger model with a topological term at $\theta=\pi$ using the Grassmann tensor renormalization group}}, \href{https://doi.org/10.1103/PhysRevD.90.074503}{\emph{Phys. Rev.} {\bfseries D90} (2014) 074503}, [\href{https://arxiv.org/abs/1408.0897}{{\ttfamily 1408.0897}}].

\bibitem{Shimizu:2017onf}
Y.~Shimizu and Y.~Kuramashi, \emph{{Berezinskii-Kosterlitz-Thouless transition in lattice Schwinger model with one flavor of Wilson fermion}}, \href{https://doi.org/10.1103/PhysRevD.97.034502}{\emph{Phys. Rev.} {\bfseries D97} (2018) 034502}, [\href{https://arxiv.org/abs/1712.07808}{{\ttfamily 1712.07808}}].

\bibitem{Butt:2019uul}
N.~Butt, S.~Catterall, Y.~Meurice, R.~Sakai and J.~Unmuth-Yockey, \emph{{Tensor network formulation of the massless Schwinger model with staggered fermions}}, \href{https://doi.org/10.1103/PhysRevD.101.094509}{\emph{Phys. Rev. D} {\bfseries 101} (2020) 094509}, [\href{https://arxiv.org/abs/1911.01285}{{\ttfamily 1911.01285}}].

\bibitem{Yosprakob:2023tyr}
A.~Yosprakob, J.~Nishimura and K.~Okunishi, \emph{{A new technique to incorporate multiple fermion flavors in tensor renormalization group method for lattice gauge theories}}, \href{https://doi.org/10.1007/JHEP11(2023)187}{\emph{JHEP} {\bfseries 11} (2023) 187}, [\href{https://arxiv.org/abs/2309.01422}{{\ttfamily 2309.01422}}].

\bibitem{Takeda:2014vwa}
S.~Takeda and Y.~Yoshimura, \emph{{Grassmann tensor renormalization group for the one-flavor lattice Gross-Neveu model with finite chemical potential}}, \href{https://doi.org/10.1093/ptep/ptv022}{\emph{PTEP} {\bfseries 2015} (2015) 043B01}, [\href{https://arxiv.org/abs/1412.7855}{{\ttfamily 1412.7855}}].

\bibitem{Akiyama:2023lvr}
S.~Akiyama, \emph{{Matrix product decomposition for two- and three-flavor Wilson fermions: Benchmark results in the lattice Gross-Neveu model at finite density}}, \href{https://doi.org/10.1103/PhysRevD.108.034514}{\emph{Phys. Rev. D} {\bfseries 108} (2023) 034514}, [\href{https://arxiv.org/abs/2304.01473}{{\ttfamily 2304.01473}}].

\bibitem{Akiyama:2020soe}
S.~Akiyama, Y.~Kuramashi, T.~Yamashita and Y.~Yoshimura, \emph{{Restoration of chiral symmetry in cold and dense Nambu--Jona-Lasinio model with tensor renormalization group}}, \href{https://doi.org/10.1007/JHEP01(2021)121}{\emph{JHEP} {\bfseries 01} (2021) 121}, [\href{https://arxiv.org/abs/2009.11583}{{\ttfamily 2009.11583}}].

\bibitem{Unmuth-Yockey:2018ugm}
J.~Unmuth-Yockey, J.~Zhang, A.~Bazavov, Y.~Meurice and S.-W. Tsai, \emph{{Universal features of the Abelian Polyakov loop in 1+1 dimensions}}, \href{https://doi.org/10.1103/PhysRevD.98.094511}{\emph{Phys. Rev.} {\bfseries D98} (2018) 094511}, [\href{https://arxiv.org/abs/1807.09186}{{\ttfamily 1807.09186}}].

\bibitem{Bazavov:2019qih}
A.~Bazavov, S.~Catterall, R.~G. Jha and J.~Unmuth-Yockey, \emph{{Tensor renormalization group study of the non-Abelian Higgs model in two dimensions}}, \href{https://doi.org/10.1103/PhysRevD.99.114507}{\emph{Phys. Rev.} {\bfseries D99} (2019) 114507}, [\href{https://arxiv.org/abs/1901.11443}{{\ttfamily 1901.11443}}].

\bibitem{Akiyama:2022eip}
S.~Akiyama and Y.~Kuramashi, \emph{{Tensor renormalization group study of (3+1)-dimensional $\mathds{Z}_{2}$ gauge-Higgs model at finite density}}, \href{https://doi.org/10.1007/JHEP05(2022)102}{\emph{JHEP} {\bfseries 05} (2022) 102}, [\href{https://arxiv.org/abs/2202.10051}{{\ttfamily 2202.10051}}].

\bibitem{Akiyama:2023hvt}
S.~Akiyama and Y.~Kuramashi, \emph{{Critical endpoint of (3+1)-dimensional finite density \ensuremath{\mathbb{Z}}$_{3}$ gauge-Higgs model with tensor renormalization group}}, \href{https://doi.org/10.1007/JHEP10(2023)077}{\emph{JHEP} {\bfseries 10} (2023) 077}, [\href{https://arxiv.org/abs/2304.07934}{{\ttfamily 2304.07934}}].

\bibitem{Fukuma:2021cni}
M.~Fukuma, D.~Kadoh and N.~Matsumoto, \emph{{Tensor network approach to two-dimensional Yang\textendash{}Mills theories}}, \href{https://doi.org/10.1093/ptep/ptab143}{\emph{PTEP} {\bfseries 2021} (2021) 123B03}, [\href{https://arxiv.org/abs/2107.14149}{{\ttfamily 2107.14149}}].

\bibitem{Hirasawa:2021qvh}
M.~Hirasawa, A.~Matsumoto, J.~Nishimura and A.~Yosprakob, \emph{{Tensor renormalization group and the volume independence in 2D U(N) and SU(N) gauge theories}}, \href{https://doi.org/10.1007/JHEP12(2021)011}{\emph{JHEP} {\bfseries 12} (2021) 011}, [\href{https://arxiv.org/abs/2110.05800}{{\ttfamily 2110.05800}}].

\bibitem{Kuwahara:2022ubg}
T.~Kuwahara and A.~Tsuchiya, \emph{{Toward tensor renormalization group study of three-dimensional non-Abelian gauge theory}}, \href{https://doi.org/10.1093/ptep/ptac103}{\emph{PTEP} {\bfseries 2022} (2022) 093B02}, [\href{https://arxiv.org/abs/2205.08883}{{\ttfamily 2205.08883}}].

\bibitem{Yosprakob:2024sfd}
A.~Yosprakob and K.~Okunishi, \emph{{Tensor renormalization group study of the three-dimensional SU(2) and SU(3) gauge theories with the reduced tensor network formulation}},  \href{https://arxiv.org/abs/2406.16763}{{\ttfamily 2406.16763}}.

\bibitem{Bloch:2022vqz}
J.~Bloch and R.~Lohmayer, \emph{{Grassmann higher-order tensor renormalization group approach for two-dimensional strong-coupling QCD}}, \href{https://doi.org/10.1016/j.nuclphysb.2022.116032}{\emph{Nucl. Phys. B} {\bfseries 986} (2023) 116032}, [\href{https://arxiv.org/abs/2206.00545}{{\ttfamily 2206.00545}}].

\bibitem{Asaduzzaman:2023pyz}
M.~Asaduzzaman, S.~Catterall, Y.~Meurice, R.~Sakai and G.~C. Toga, \emph{{Tensor network representation of non-abelian gauge theory coupled to reduced staggered fermions}}, \href{https://doi.org/10.1007/JHEP05(2024)195}{\emph{JHEP} {\bfseries 05} (2024) 195}, [\href{https://arxiv.org/abs/2312.16167}{{\ttfamily 2312.16167}}].

\bibitem{Kuhn:2015zqa}
S.~K\"uhn, E.~Zohar, J.~I. Cirac and M.~C. Ba\~nuls, \emph{{Non-Abelian string breaking phenomena with Matrix Product States}}, \href{https://doi.org/10.1007/JHEP07(2015)130}{\emph{JHEP} {\bfseries 07} (2015) 130}, [\href{https://arxiv.org/abs/1505.04441}{{\ttfamily 1505.04441}}].

\bibitem{Silvi:2016cas}
P.~Silvi, E.~Rico, M.~Dalmonte, F.~Tschirsich and S.~Montangero, \emph{{Finite-density phase diagram of a (1+1)-d non-abelian lattice gauge theory with tensor networks}}, \href{https://doi.org/10.22331/q-2017-04-25-9}{\emph{Quantum} {\bfseries 1} (2017) 9}, [\href{https://arxiv.org/abs/1606.05510}{{\ttfamily 1606.05510}}].

\bibitem{Banuls:2017ena}
M.~C. Ba\~nuls, K.~Cichy, J.~I. Cirac, K.~Jansen and S.~K\"uhn, \emph{{Efficient basis formulation for 1+1 dimensional SU(2) lattice gauge theory: Spectral calculations with matrix product states}}, \href{https://doi.org/10.1103/PhysRevX.7.041046}{\emph{Phys. Rev. X} {\bfseries 7} (2017) 041046}, [\href{https://arxiv.org/abs/1707.06434}{{\ttfamily 1707.06434}}].

\bibitem{Sala:2018dui}
P.~Sala, T.~Shi, S.~K\"uhn, M.~C. Ba\~nuls, E.~Demler and J.~I. Cirac, \emph{{Variational study of U(1) and SU(2) lattice gauge theories with Gaussian states in 1+1 dimensions}}, \href{https://doi.org/10.1103/PhysRevD.98.034505}{\emph{Phys. Rev. D} {\bfseries 98} (2018) 034505}, [\href{https://arxiv.org/abs/1805.05190}{{\ttfamily 1805.05190}}].

\bibitem{Silvi:2019wnf}
P.~Silvi, Y.~Sauer, F.~Tschirsich and S.~Montangero, \emph{{Tensor network simulation of an SU(3) lattice gauge theory in 1D}}, \href{https://doi.org/10.1103/PhysRevD.100.074512}{\emph{Phys. Rev. D} {\bfseries 100} (2019) 074512}, [\href{https://arxiv.org/abs/1901.04403}{{\ttfamily 1901.04403}}].

\bibitem{Rigobello:2023ype}
M.~Rigobello, G.~Magnifico, P.~Silvi and S.~Montangero, \emph{{Hadrons in (1+1)D Hamiltonian hardcore lattice QCD}},  \href{https://arxiv.org/abs/2308.04488}{{\ttfamily 2308.04488}}.

\bibitem{Liu:2023lsr}
H.~Liu, T.~Bhattacharya, S.~Chandrasekharan and R.~Gupta, \emph{{Phases of 2d massless QCD with qubit regularization}},  \href{https://arxiv.org/abs/2312.17734}{{\ttfamily 2312.17734}}.

\bibitem{Hayata:2023pkw}
T.~Hayata, Y.~Hidaka and K.~Nishimura, \emph{{Dense QCD$_{2}$ with matrix product states}}, \href{https://doi.org/10.1007/JHEP07(2024)106}{\emph{JHEP} {\bfseries 07} (2024) 106}, [\href{https://arxiv.org/abs/2311.11643}{{\ttfamily 2311.11643}}].

\bibitem{Kogut:1983ia}
J.~B. Kogut, H.~Matsuoka, M.~Stone, H.~W. Wyld, S.~H. Shenker, J.~Shigemitsu et~al., \emph{{Chiral Symmetry Restoration in Baryon Rich Environments}}, \href{https://doi.org/10.1016/0550-3213(83)90014-7}{\emph{Nucl. Phys. B} {\bfseries 225} (1983) 93--122}.

\bibitem{NAKAMURA1984391}
A.~Nakamura, \emph{Behavior of quarks and gluons at finite temperature and density in $su(2)$ qcd}, \href{https://doi.org/https://doi.org/10.1016/0370-2693(84)90430-1}{\emph{Phys. Lett. B} {\bfseries 149} (1984) 391--395}.

\bibitem{Akiyama:2022pse}
S.~Akiyama, \emph{{Bond-weighting method for the Grassmann tensor renormalization group}}, \href{https://doi.org/10.1007/JHEP11(2022)030}{\emph{JHEP} {\bfseries 11} (2022) 030}, [\href{https://arxiv.org/abs/2208.03227}{{\ttfamily 2208.03227}}].

\bibitem{Hands:1999md}
S.~Hands, J.~B. Kogut, M.-P. Lombardo and S.~E. Morrison, \emph{{Symmetries and spectrum of SU(2) lattice gauge theory at finite chemical potential}}, \href{https://doi.org/10.1016/S0550-3213(99)00364-8}{\emph{Nucl. Phys. B} {\bfseries 558} (1999) 327--346}, [\href{https://arxiv.org/abs/hep-lat/9902034}{{\ttfamily hep-lat/9902034}}].

\bibitem{Kogut:1999iv}
J.~B. Kogut, M.~A. Stephanov and D.~Toublan, \emph{{On two color QCD with baryon chemical potential}}, \href{https://doi.org/10.1016/S0370-2693(99)00971-5}{\emph{Phys. Lett. B} {\bfseries 464} (1999) 183--191}, [\href{https://arxiv.org/abs/hep-ph/9906346}{{\ttfamily hep-ph/9906346}}].

\bibitem{Kogut:2002cm}
J.~B. Kogut, D.~Toublan and D.~K. Sinclair, \emph{{The Phase diagram of four flavor SU(2) lattice gauge theory at nonzero chemical potential and temperature}}, \href{https://doi.org/10.1016/S0550-3213(02)00678-8}{\emph{Nucl. Phys. B} {\bfseries 642} (2002) 181--209}, [\href{https://arxiv.org/abs/hep-lat/0205019}{{\ttfamily hep-lat/0205019}}].

\bibitem{Nishida:2003uj}
Y.~Nishida, K.~Fukushima and T.~Hatsuda, \emph{{Thermodynamics of strong coupling two color QCD with chiral and diquark condensates}}, \href{https://doi.org/10.1016/j.physrep.2004.05.005}{\emph{Phys. Rept.} {\bfseries 398} (2004) 281--300}, [\href{https://arxiv.org/abs/hep-ph/0306066}{{\ttfamily hep-ph/0306066}}].

\bibitem{Hands:2007uc}
S.~Hands, P.~Sitch and J.-I. Skullerud, \emph{{Hadron Spectrum in a Two-Colour Baryon-Rich Medium}}, \href{https://doi.org/10.1016/j.physletb.2008.01.078}{\emph{Phys. Lett. B} {\bfseries 662} (2008) 405--412}, [\href{https://arxiv.org/abs/0710.1966}{{\ttfamily 0710.1966}}].

\bibitem{Strodthoff:2011tz}
N.~Strodthoff, B.-J. Schaefer and L.~von Smekal, \emph{{Quark-meson-diquark model for two-color QCD}}, \href{https://doi.org/10.1103/PhysRevD.85.074007}{\emph{Phys. Rev. D} {\bfseries 85} (2012) 074007}, [\href{https://arxiv.org/abs/1112.5401}{{\ttfamily 1112.5401}}].

\bibitem{Cotter:2012mb}
S.~Cotter, P.~Giudice, S.~Hands and J.-I. Skullerud, \emph{{Towards the phase diagram of dense two-color matter}}, \href{https://doi.org/10.1103/PhysRevD.87.034507}{\emph{Phys. Rev. D} {\bfseries 87} (2013) 034507}, [\href{https://arxiv.org/abs/1210.4496}{{\ttfamily 1210.4496}}].

\bibitem{Braguta:2016cpw}
V.~V. Braguta, E.~M. Ilgenfritz, A.~Y. Kotov, A.~V. Molochkov and A.~A. Nikolaev, \emph{{Study of the phase diagram of dense two-color QCD within lattice simulation}}, \href{https://doi.org/10.1103/PhysRevD.94.114510}{\emph{Phys. Rev. D} {\bfseries 94} (2016) 114510}, [\href{https://arxiv.org/abs/1605.04090}{{\ttfamily 1605.04090}}].

\bibitem{Iida:2019rah}
K.~Iida, E.~Itou and T.-G. Lee, \emph{{Two-colour QCD phases and the topology at low temperature and high density}}, \href{https://doi.org/10.1007/JHEP01(2020)181}{\emph{JHEP} {\bfseries 01} (2020) 181}, [\href{https://arxiv.org/abs/1910.07872}{{\ttfamily 1910.07872}}].

\bibitem{Iida:2024irv}
K.~Iida, E.~Itou, K.~Murakami and D.~Suenaga, \emph{{Lattice study on finite density QC$_2$D towards zero temperature}},  \href{https://arxiv.org/abs/2405.20566}{{\ttfamily 2405.20566}}.

\bibitem{Mermin:1966fe}
N.~D. Mermin and H.~Wagner, \emph{{Absence of ferromagnetism or antiferromagnetism in one-dimensional or two-dimensional isotropic Heisenberg models}}, \href{https://doi.org/10.1103/PhysRevLett.17.1133}{\emph{Phys. Rev. Lett.} {\bfseries 17} (1966) 1133--1136}.

\bibitem{Coleman:1973ci}
S.~R. Coleman, \emph{{There are no Goldstone bosons in two-dimensions}}, \href{https://doi.org/10.1007/BF01646487}{\emph{Commun. Math. Phys.} {\bfseries 31} (1973) 259--264}.

\bibitem{Yoshimura:2017jpk}
Y.~Yoshimura, Y.~Kuramashi, Y.~Nakamura, S.~Takeda and R.~Sakai, \emph{{Calculation of fermionic Green functions with Grassmann higher-order tensor renormalization group}}, \href{https://doi.org/10.1103/PhysRevD.97.054511}{\emph{Phys. Rev.} {\bfseries D97} (2018) 054511}, [\href{https://arxiv.org/abs/1711.08121}{{\ttfamily 1711.08121}}].

\bibitem{MORITA201965}
S.~Morita and N.~Kawashima, \emph{Calculation of higher-order moments by higher-order tensor renormalization group}, \href{https://doi.org/https://doi.org/10.1016/j.cpc.2018.10.014}{\emph{Computer Physics Communications} {\bfseries 236} (2019) 65 -- 71}.

\bibitem{Cohen:2003kd}
T.~D. Cohen, \emph{{Functional integrals for QCD at nonzero chemical potential and zero density}}, \href{https://doi.org/10.1103/PhysRevLett.91.222001}{\emph{Phys. Rev. Lett.} {\bfseries 91} (2003) 222001}, [\href{https://arxiv.org/abs/hep-ph/0307089}{{\ttfamily hep-ph/0307089}}].

\bibitem{Schon:2000he}
V.~Schon and M.~Thies, \emph{{Emergence of Skyrme crystal in Gross-Neveu and 't Hooft models at finite density}}, \href{https://doi.org/10.1103/PhysRevD.62.096002}{\emph{Phys. Rev. D} {\bfseries 62} (2000) 096002}, [\href{https://arxiv.org/abs/hep-th/0003195}{{\ttfamily hep-th/0003195}}].

\bibitem{Thies:2003kk}
M.~Thies and K.~Urlichs, \emph{{Revised phase diagram of the Gross-Neveu model}}, \href{https://doi.org/10.1103/PhysRevD.67.125015}{\emph{Phys. Rev. D} {\bfseries 67} (2003) 125015}, [\href{https://arxiv.org/abs/hep-th/0302092}{{\ttfamily hep-th/0302092}}].

\bibitem{Kojo:2011fh}
T.~Kojo, \emph{{A (1+1) dimensional example of Quarkyonic matter}}, \href{https://doi.org/10.1016/j.nuclphysa.2011.12.002}{\emph{Nucl. Phys. A} {\bfseries 877} (2012) 70--94}, [\href{https://arxiv.org/abs/1106.2187}{{\ttfamily 1106.2187}}].

\bibitem{qsw}
\url{https://qsw.phys.s.u-tokyo.ac.jp/}.

\end{thebibliography}\endgroup

\end{document}